\documentclass[10pt,final,twocolumn]{IEEEtran}
\usepackage{ amsmath,graphicx}
\usepackage{tabularx,booktabs}
\newcolumntype{C}{>{\centering\arraybackslash}X}
\usepackage{lipsum}
\usepackage{algorithmicx}
\usepackage{algorithm}
\usepackage{algpseudocode}
\usepackage{amsfonts}
\usepackage{amsmath}
\usepackage{amssymb}
\usepackage{bigstrut}
\usepackage{bbm}
\usepackage{bm}
\usepackage{caption}
\usepackage{cite}
\usepackage{color}
\usepackage{colortbl}
\usepackage{eqparbox}
\usepackage{epstopdf}
\usepackage{epsfig}
\usepackage{epsf}
\usepackage{float}
\usepackage{footnote}
\usepackage{graphicx}
\usepackage{latexsym}
\usepackage{mwe}
\usepackage{mdwmath}
\usepackage{mdwtab}
\usepackage{placeins}
\usepackage{rotating}
\usepackage{stackengine}
\usepackage{subcaption}
\usepackage{syntonly}
\usepackage{tabularx}
\usepackage{times}
\usepackage{url}
\usepackage{verbatim}
\usepackage{wrapfig}
\usepackage{xspace}
\usepackage{epstopdf}
\usepackage{algpseudocode}
\usepackage{algorithmicx}
\usepackage{algorithm}

\def \bW {{\mathbf W}}
\def \bB {{\mathbf B}}
\def \bA {{\mathbf A}}
\def \bC {{\mathbf C}}
\def \bH {{\mathbf H}}
\def \bW {{\mathbf W}}
\def \bZ {{\mathbf Z}}
\def \bY {{\mathbf Y}}
\def \ba {{\mathbf a}}
\def \bb {{\mathbf b}}
\def \bc {{\mathbf c}}
\def \bV {{\mathbf V}}
\def \bU {{\mathbf U}}
\def \bu {{\mathbf u}}
\def \bD {{\mathbf D}}

\begin{document}
%

\title{Identification of Overlapping Communities via\\ Constrained Egonet Tensor Decomposition}
%

%


\author{\textit{Fatemeh Sheikholeslami and Georgios B. Giannakis}\\
	Dept. of ECE and Digital Tech. Center, University of Minnesota\\
	Minneapolis, MN 55455, USA\\
	E-mails: \{sheik081,georgios\}@umn.edu
	\thanks{Work in this paper was supported by NSF grants  1500713, 1442686, 1514056, and NIH grant no.  1R01GM104975-01.}}

\maketitle
\begin{abstract}
Detection of overlapping communities in real-world networks is a generally challenging task. Upon recognizing that a network is in fact the union of its egonets,  a novel network representation using multi-way data structures is advocated in this contribution. The introduced sparse tensor-based representation exhibits  richer structure compared to its matrix counterpart, and thus enables  a more robust approach to community detection. To leverage this structure, a constrained tensor approximation framework is introduced using PARAFAC decomposition.  The arising constrained trilinear optimization is handled via alternating minimization, where intermediate subproblems are solved  using the alternating direction method of multipliers (ADMM) to ensure convergence. The   factors obtained provide soft community memberships, which can further be  exploited for crisp, and possibly-overlapping community assignments. The framework is further broadened to include  time-varying graphs, where the edgeset as well as the underlying communities evolve through time. Performance of the proposed approach is assessed via tests on benchmark synthetic graphs as well as real-world networks. As corroborated by numerical tests, the proposed tensor-based  representation captures multi-hop nodal connections, that is, connectivity patterns within single-hop neighbors, whose exploitation yields a more robust community identification in the presence of mixing as well as  overlapping communities.
\end{abstract}

\section{Introduction}
\label{sec:introduction}
Graph representation of complex real-world networks provides an invaluable tool for analysis and discovery of intrinsic attributes present in social, biological, and financial networks. One such attribute is the presence of small subgraphs, referred to as ``communities'' or ``clusters,'' whose dense intra-connections and sparse inter-connections often represents a potential ``association'' among the participating entities (nodes). The task of community identification targets the discovery of such  highly-interwoven nodes, and is of paramount interest in areas as diverse as unveiling functional modules in biological networks such as brain~\cite{brain1}, trend analysis in social media~\cite{social,social2}, and clustering of costumers in recommender systems~\cite{recsys}.

Past works on community detection include those based on modularity-maximization~\cite{louvain,extremal}, generative and statistical models~\cite{MMSB, kakade,bigclam}, local-metric optimization\cite{CPM}, spectral clustering~\cite{spectralclustering}, and matrix factorization~\cite{NMF_kdd2011,NMF_nature2013,BayesianNMF_2011,trilinear_NMF_kdd2012,symmetricNMF_phy2013}; see e.g. ~\cite{Fortu,fortunato2016userguid}  for a comprehensive overview. With recent exploratoty studies over contemporary real-world networks, new challenges have been raised in  community identification, addressing the presence of overlapping communities ~\cite{2013overlapping,ICDM2015overlapping,overlapping_k_means_ICDM2015}, multimodal interaction of nodes  over multiview networks~\cite{papalexakis2013Fusion,papalexakis}, exploitation of nodal and edge-related side-information~\cite{ICDM2013_Nodal}, as well as dynamic interactions within a network~\cite{comet2014,baingana2016joint}.

In handling such new challenges, reliance on the adjacency matrix representation of networks limits their capabilities in capturing higher-order interactions, which can potentially provide critical information via temporal, multi-modal, or even multi-hop connectivity among nodes. To this end, \textit{tensors } as multi-way data structures provide a  viable alternative, whose increased representational capabilities can potentially lead to a more informative community identification \cite{papalexakis2013Fusion,papalexakis,comet2014,jmlr2015_mmsb,globalsip2016}.
For instance,~\cite{tensor_lescovec} and \cite{tophits_ICDM2005} construct higher-order tensors whose entries
are non-zero if a tuple of nodes jointly belongs to a cycle or a clique, while \cite{comet2014} captures temporal dynamics of communities via tensors.
Under certain conditions, tensor decompositions are unique
~\cite{Kru77,SidBro00,siam_tensor}, and can guarantee identifiability of the community structure.

The present work develops a novel tensor-based network representation by recognizing that a network is the union of its \textit{egonets}. An egonet is defined per node as the subgraph induced by the node itself, its one-hope neighbors, and all their connections, whose structure has been exploited in anomaly detection~\cite{oddball}, and user-specific community identification ~\cite{leskovec_nips2012_ego,socialego_2016}. By concatenating egonet adjacency matrices along the  $3$-rd dimension  of a three-way tensor, the proposed network representation, named \textit{egonet-tensor},  captures  information per node beyond its one-hop connectivity patterns. In fact, in a number of practical networks only adjacency  matrix of the network is given, rendering egonets a unique candidate for enhancing community identification performance when extra nodal features are kept private, e.g., Amazon costumer graphs.  By construction, egonet-tensor exhibits richer structure compared to its matrix counterpart, which is further exploited by casting the community detection task in   a constrained tensor decomposition framework.  Building on preliminary results in \cite{globalsip2016}, solvers with convergence guarantees are developed for the proposed constrained non-convex optimization, whose solution yields the community-revealing components,  utilized for soft as well as crisp community assignments unveiling possibly \textit{overlapping} communities.
The performance of the proposed EgoTen toolbox is compared with its matrix counterpart as well as state-of-the-art methods, corroborating the improved quality of detected communities through the exploitation of higher-order statistics captured  in the  egonet-tensor representations.  Furthermore, the proposed tensor-based framework is extended for application on time-varying graphs, where network connectivity evolves through time, leading to emergence or disappearance of communities.

The rest of the paper is organized as follows. Section \ref{sec:preliminaries} introduces the novel tensor-based network representation, and  Section \ref{sec:cpd} presents the  constrained tensor decomposition, and its efficient solver  for the task of community identification. Performance metrics for evaluating the quality of detected communities in networks with and without ground-truth communities is the subject of Section \ref{sec:metrics}, and Section \ref{sec:metrics} introduces identification of time-varying graphs using the proposed algorithms.  Section \ref{sec:tests} provides numerical tests, while Section \ref{sec:conclusion} concludes the paper.

\emph{Notation.} Lower- (upper-) case boldface letters denote column vectors (matrices), and underlined upper-case boldface letters stand for tensor structures. Calligraphic symbols are reserved for sets, while $^T$ stands for transposition. Symbols $\circ$, $\otimes$ and  $\odot$ are reserved for outer-product, Kronecker-product and Khatri-Rao-product, respectively, and $\mathrm{Tr\{\mathbf{X}\}}$ denotes the trace of matrix $\mathbf{X}$.

\section{Egonet-Tensor Construction}
\label{sec:preliminaries}
Consider a graph  $\mathcal{G} = (\mathcal{V},\mathcal{E}, \mathbf{W})$, where
$\mathcal{V}$ , $\mathcal{E}$, and $\mathbf{W} \in \mathbb{R}^{N \times N}$
respectively denote the set of $N$ nodes, i.e., $|\mathcal{V}|=N$, edges, and the adjacency matrix. In the case of binary networks, $w_{ij} :=1 $
 if $(i,j) \in \mathcal{E}$, and  $w_{ij} := 0$ otherwise.
Furthermore, the \emph{egonet of node $n$} is defined as the subgraph induced by node $n$, its single-hop neighbors, and all their connections. Let  $\mathcal{G}^{(n)}:=(\mathcal{V},\mathcal{E}^{(n)}, \mathbf{W}^{(n)}) \subset \mathcal{G}$ be the subgraph with  $\mathcal{E}^{(n)}$ the egonet edgeset, and $\mathbf{W}^{(n)} \in \mathbb{R}^{N \times N}$  the corresponding adjacency matrix whose non-zero support captures the edges in $\mathcal{E}^{(n)}$; that is,

\[
w_{ij}^{(n)} :=
\begin{cases}
w_{ij} & \text{if } (i,j) \in \mathcal{E}^{(n)} \\
0 & \text{otherwise.}
\end{cases}
\]
Figure 1 (a) illustrates  such subgraphs, where the black node corresponds to the central node of the egonet, and the single-hop neighbors are colored green. Typically, the center node $n$ is  excluded from $\mathcal{G}^{(n)}$, but it is included here for convenience.

As Figure 1 (b) depicts, graph $\mathcal{G}$ can now be fully described by a three-way \emph{egonet-tensor} $\underline{\mathbf{W}} \in \mathbb{R}^{N \times N \times N}$, where frontal slabs correspond to egonet adjacency matrices $\{\mathbf{W}^{(n)}\}_{n=1}^N$, stacked one after the other. In tensor parlance, that is tantamount to setting the $n$-th frontal slab as $\underline{\mathbf{W}}_{:,:,n}:= \mathbf{W}^{(n)}$, where $:$ is a free index that spans its range. 

\begin{figure}[t]
	
	\begin{minipage}[b]{1\linewidth}

		\centering
		\centerline{\includegraphics[width=0.8\textwidth]{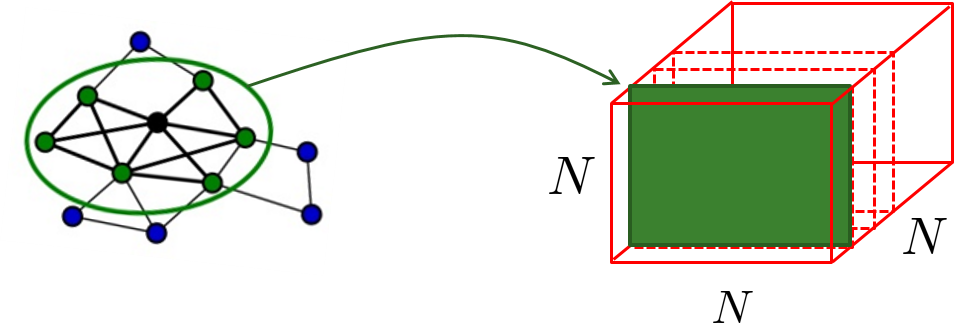}}
		\centerline{(a) \hspace{4cm}(b)}\medskip
		\caption{Construction of \emph{egonet-tensor} with frontal slabs as egonet adjacencies.}
		\label{fig:toy}
	\end{minipage}
\end{figure}

The advantage of presenting a graph with its egonet-tensor lies in the fact that tensors as higher-order structures are capable of capturing {``useful information''} along their different dimensions, a.k.a., ``modes.'' In a network with underlying community structure, the proposed egonet-tensor representation is indeed capable of preserving the inherent ``similarities'' among egonet adjacencies along the $3$-rd mode, whose extraction is crucial in tasks such as community detection. Such representation is of particular interest  for various settings where no nodal features are provided (that is, only the adjacency matrix is given), making egonets very appealing  for an improved network representation as well as the development of robust schemes for the task of interest.
 The following toy example clarifies how such similarities induce a structure over the egonet tensor, and intuitively discusses how its exploitation can lead to an improved performance. 

\subsection{Toy example}

Let us consider a toy network with three fully-connected communities, illustrated  in Figure~\ref{fig:toy} (a). Since each node is a member of a \textit{fully-connected} community, the binary adjacency matrix of its egonet is identical to that of any other node in its resident community. Furthermore, after permutation, this egonet adjacency matrix consists of a single block of nonzero  entries (with zero diagonal entries if the network is free of self-loops), and zeros elsewhere. This implies that the egonet-tensor can be  permuted similarly into a block-diagonal tensor with three nonzero blocks, as illustrated in Figure~\ref{fig:toy} (b).

\begin{figure}[]
	\begin{minipage}[b]{0.5\linewidth}
		\centering
		\centerline{\includegraphics[width=0.5\textwidth]{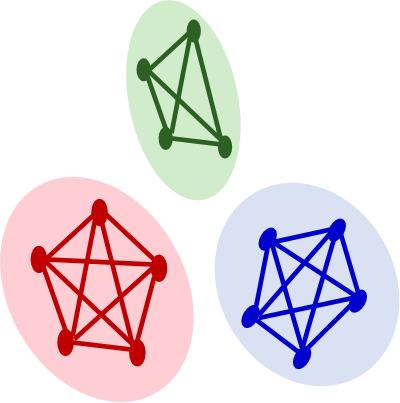}}
		\centerline{(a) }\medskip
	\end{minipage}
	\begin{minipage}[b]{.48\linewidth}
		\centering
		\centerline{\includegraphics[width=0.5\textwidth]{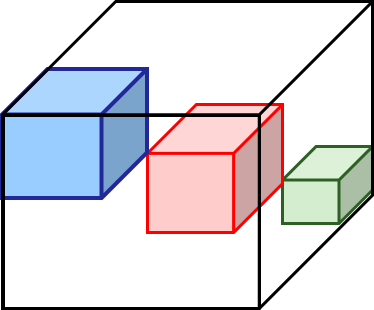}}
		\centerline{(b) }\medskip
	\end{minipage}
	\hfill
	
	\begin{minipage}[b]{1\linewidth}
		\centering
		\centerline{\includegraphics[width=0.73\textwidth]{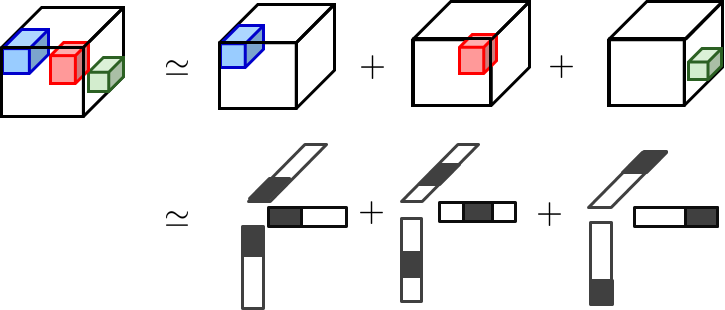}}
		\centerline{(c) }\medskip
	\end{minipage}

	\caption{(a) A toy  network with $3$ fully connected non-overlapping communities;
		(b) corresponding egonet-tensor; and (c) its community-revealing factorization via  CPD.}
	\label{fig:toy2}
\end{figure}

Adopting the well-known canonical polyadic decomposition (CPD) to decompose $\underline{\mathbf{W}}$ into its constituent rank-one tensors, the model naturally approximates the tensor with rank three, thus revealing the number of underlying communities. In fact, if the diagonal entries were all set to 1, i.e. considering self loops for all the neighboring nodes in an egonet, this approximation would be exact; see Figure~\ref{fig:toy} (c).

In practice, real-world networks often demonstrate overlapping community structure, where some nodes are associated with multiple communities rather than a single one. To address such cases, consider the augmented network in Figure 3 (a), where a new node (or a super node corresponding to more than one node) associated with two communities is added. Once the corresponding egonet tensor is constructed, the presence of overlapping communities manifests itself in overlapping diagonal blocks in the egonet tensor; see Figure 3 (b) in comparison with disjoint blocks in Figure 2 (b). As it will become evident in Section II, by exploiting a strutured CPD on the arising egonet-tensor, it can be shown that the  frontal slab corresponding to the egonet adjacency of an overlapping node is approximated by multiple summands, each  corresponding to one of its resident communities. Figure 3 (c) illustrates how the obtained decomposition reveals the multi-community association of the augmented node based on its egonet adjacency matrix. The egonet-tensor representation naturally trades off flexibility for increased redundancy and memory costs. Nevertheless, the resulting tensor is extremely sparse, and off-the-shelf tools for sparse tensor computations can be readily utilized; see e.g.,  \cite{kolda,papalexakis2012parcube,splatt}.

Unfortunately, such idealistic assumptions where each community is fully-connected within and well-separated from other communities, are not fulfilled in real-world networks. Nevertheless, the inherent similarities among egonet adjacencies induce a potentially useful reinforced structure along the $3^{\text{rd}}$ mode of the proposed egonet-tensor representation. 
For instance,  nodes in a community often exhibit dense (rather than full) connections  among themselves, and fewer connections with other communities. This property is consequently reflected in the egonet-tensor (as well as the traditional matrix adjacency) representation  by dense diagonal blocks, whose clear separation fades aways as\textit{ out-of-community connections} increase. However, the presence of overlapping communities can further smear the block-structure as overlapping nodes are ``{ well-connected}'' with multiple communities. The performance of  traditional community detection methods often dramatically degrades in networks with such properties, whereas exploiting the {\it structured redundancy} offered via the proposed egonet-tensor representation and casting the problem in a comunity-revealing tensor decomposition framework increases robustness against the aforementioned phenomena.

\begin{figure}[]
	\begin{minipage}[b]{0.5\linewidth}
		\centering
		\centerline{\includegraphics[width=0.6\textwidth]{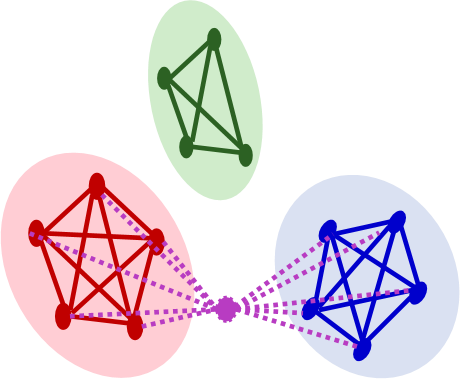}}
		\centerline{(a) }\medskip
	\end{minipage}
	\begin{minipage}[b]{.4\linewidth}
		\centering
		\centerline{\includegraphics[width=0.6\textwidth]{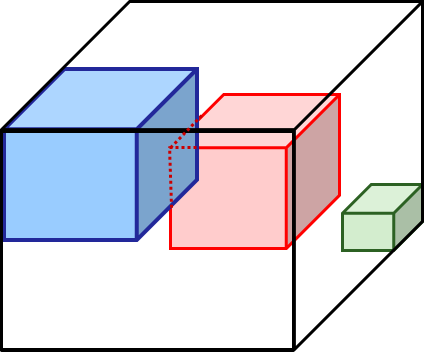}}
		\centerline{(c) }\medskip
	\end{minipage}
	\begin{minipage}[b]{1\linewidth}
		\centering
		\centerline{\includegraphics[width=0.7\textwidth]{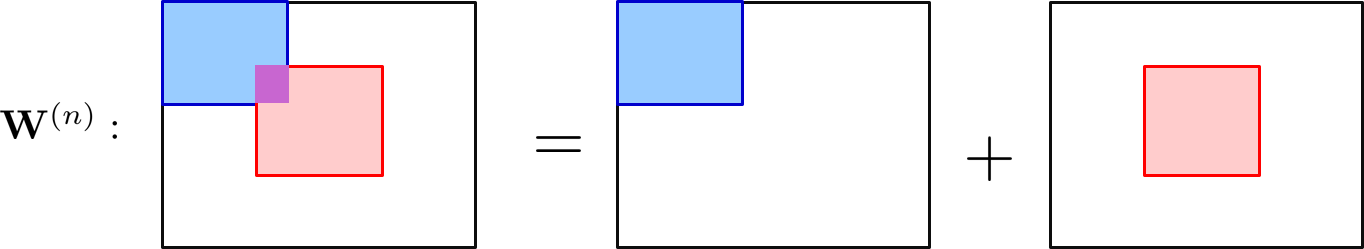}}
		\centerline{(b) }\medskip
	\end{minipage}
	\hfill
	
	\caption{(a) A toy  network with overlapping communities;
		(b) corresponding egonet-tensor; and (c) community-revealing approximation of egonet adjacency of the shared node via  CPD.}
	\label{fig:toy}
\end{figure}


\section{Constrained Tensor Decomposition}
\label{sec:cpd}

Given $\underline{\mathbf{W}}$, this section leverages the canonical polyadic decomposition (CPD) \cite{kolda} in order to factorize the egonet tensor into its constituent community-revealing factors. Assuming that the number of communities is upperbounded by $K$, a rank-$K$ CPD model is sought by solving the following constrained least-squares (LS) problem 
\begin{eqnarray}
\label{parafac}
\nonumber
\{ \widehat{\bA}, \widehat{\bB}, \widehat{\bC} \} = \underset{\bA,\bB,\bC}{\text{arg min}} &
\|\underline{\bW} - \sum_{k=1}^K \ba_k \circ \bb_k\circ \bc_k  \|_F^2 \\
\text{s.t.} &   \bA\geq \mathbf{0},\bB\geq \mathbf{0},\bC \geq \mathbf{0}
\end{eqnarray}
where $\bA :=[\ba_1,...,\ba_K]\in \mathbb{R}^{N \times K}$, $\bB:=[\bb_1,...,\bb_K] \in \mathbb{R}^{N \times K}$, and $\bC:=[\bc_1,...,\bc_K] \in \mathbb{R}^{N \times K}$; while the term ($\ba_k \circ \bb_k \circ \bc_k$) is the outer product of the three vectors, which induces the $k$-th rank-one tensor component in the rank-$K$ decomposition; see \cite{kolda} for further details on CPD. The constraint $\bA \geq \mathbf{0}$ denotes entry-wise nonnegativity constraints, i.e., $a_{nk} \geq 0$ for $n=1,\ldots, N$ and $k=1,\ldots,K$; and similarly for factors $\bB$ and $\bC$. These constraints enforce the nonnegativity of  egonet adjacency matrices, thus inducing structure in the sought CPD  and providing interpretation of the decomposition factors. 

It is possible to re-write \eqref{parafac} as, see e.g., \cite{kolda}
\begin{eqnarray*}
\label{parafac2}
\nonumber
\{ \widehat{\bA}, \widehat{\bB}, \widehat{\bC} \} = \underset{\bA,\bB,\bC}{\text{arg min}} &
\sum_{n=1}^N \|{\bW}^{(n)} -  \bA \text{diag}(\tilde{\bc}_n) \bB^\top \|_F^2 \\
\text{s.t.} &   \bA\geq \mathbf{0},\bB\geq \mathbf{0},\bC \geq \mathbf{0}
\end{eqnarray*}
where $\text{diag}(\tilde{c}_n)$ is a diagonal matrix holding the $n$-th row of $\bC$ on its diagonal. %
Focusing on the $n$-th frontal slab of the egonet-tensor, CPD provides the approximation 
\begin{equation}
\mathbf{W}^{(n)} = \sum_{k=1}^K c_{nk} (\ba_k  \bb_k^\top)
\end{equation}
where $c_{nk}$ denotes the $(n,k)$-th entry of factor $\bC$. Such decomposition can be interpreted as a weighted sum over $K$ ``basis'', $\{\ba_k  \bb_k^\top\}_{k=1}^K$, where $(\ba_k  \bb_k^\top)$ captures the ``connectivity structure'' within the $k$-th community. Consequently, $c_{nk}$ can be viewed as \textit{association level} of node $n$ to community $k$ for $k=1,...,K$. Furthermore,  one can easily realize that since $(\ba_k  \bb_k^\top)$ is viewed as connectivity structure of community $k$, the elements in $\ba_k:=[a_{1k},...,a_{nk}]^\top$ and $\bb_k:=[b_{1k},...,b_{nk}]^\top$ can be consequently viewed as the \emph{contribution levels} of nodes $n=1,...,N$ to community $k$. 
This interpretation of the factors, prompts us to further leverage the structure and introduce additional constraints on the CPD factors.

\subsection{Structured CPD}
Real-world networks often involve nodes which are associated with more than one community, resulting in multiple nonzero entries in the association vector $[c_{n1},c_{n2},...,c_{nK}]$ corresponding to a generic node $n$.  
To return a normalized association vector, we augment the optimization in \eqref{parafac} by the constraints $\sum_{k=1}^K c_{nk}=1$ for $n=1,2,...,N$, which together with $\bC \geq \textbf{0}$ acts as a simplex constraint on the rows of $\bC$. Upon imposing simplex constraints the CPD is further regularized as 
\begin{eqnarray}
\label{parafac2}
\nonumber
 \{ \widehat{\bA}, \widehat{\bB}, \widehat{\bC} \}   = &\underset{\bA,\bB,\bC}{\text{arg min}} 
\Big{\{}\|\underline{\bW} - \sum_{k=1}^K \ba_k \circ \bb_k\circ \bc_k  \|_F^2 \\ 
&+ \lambda(\|\bA\|_F^2+\|\bB\|_F^2)\Big{\}}\\
& \hspace{-1cm}\text{s.t.}  \qquad   \bA\geq \mathbf{0},\bB\geq \mathbf{0},\bC \geq \mathbf{0}\nonumber  \\
 & \qquad \qquad \|\tilde{\bc}_n\|_1=1  \qquad  \forall n=1,2,...,N \nonumber \nonumber
\end{eqnarray}
 Different from \cite{morteza} and \cite{juan}, the regularization term $\|\bA\|_F^2+\|\bB\|_F^2$ does not play the role of rank-regularization for subspace learning, instead it solves the scalar ambiguity in factors $\bA$ and $\bB$.

The CPD problem formulated in \eqref{parafac2} is a tri-linear constrained LS problem, whose minimization can be tackled by alternating optimization. In the ensuing subsection, the proposed solver is developed using by  alternating optimization with ADMM intermediate steps; see e.g. \cite{admm_tutor} and \cite{keijun}.

\subsection{Solving the proposed structured CPD} 
 In the proposed alternating optimization scheme, each step consists of fixing two factors and minimizing the arising subproblem with respect to the third factor. In this subsection, we study the emerging sub-problems and propose efficient solvers for tackling those. 
\subsubsection{Factor $\bA$ update}
Consider first the update of factor $\bA$ at iteration $k$, obtained after fixing $\bB = \bB^{(k-1)}$ and $\bC = \bC^{(k-1)}$ and solving the corresponding minimization. The arising subproblem,  after algebric manipulation can be re-written as
\begin{equation}\label{Aupdate}
 {\bA}^{(k)}   = \underset{\bA\geq \mathbf{0}}{\text{arg min}} 
\| {\bW_1} - (\bB^{(k-1)} \odot \bC^{(k-1)}) \bA ^\top \|_F^2 + \lambda \|\bA\|_F^2
\end{equation}
where $\bW_1:= [\text{vec}(\underline{\bW}_{1,:,:}) , \ldots , \text{vec}(\underline{\bW}_{N,:,:})] \in \mathbf{R}^{N^2 \times N}$ is a matricized reshaping of the tensor $\underline{\mathbf{W}}$. Also $\mathbf{B}^{(k-1)} \odot \mathbf{C}^{(k-1)} := \left[ \mathbf{b}^{(k-1)}_1 \otimes \mathbf{c}^{(k-1)}_1, \ldots, \mathbf{b}^{(k-1)}_K \otimes \mathbf{c}^{(k-1)}_K  \right]$ is the Khatri-Rao product of $\mathbf{B}^{(k-1)}$ and $\mathbf{C}^{(k-1)}$, where $\mathbf{b}^{(k-1)}_i$  $(\mathbf{c}^{(k-1)}_i)$ denotes column $i$ of $\bB^{(k-1)}$ (resp. $\bC^{(k-1)}$), and $\otimes$ denotes the Kronecker product operator; see also~\cite{kolda}.  
%

 Following the steps in \cite{keijun}, auxiliary variable $\bar{\bA}$ is introduced to account for the nonnegativity constraint, and the  augmented Lagrangian of \eqref{Aupdate} is
\begin{eqnarray}
  \mathcal{L}^{(k)}_A(\bA,\bar{\bA},Y)
  =& \| {\bW_1} - \bH_A^{(k)}\bA ^\top \|_F^2 + \lambda \text{Tr}\{\mathbf{A}  \bA^\top \}\nonumber \\ &   + r_{+}(\bar{\bA}) + (\rho/2) \|\bY+\bA-\bar{\bA}\|_F^2 
\end{eqnarray}
where $\bar{\mathbf{A}},\mathbf{Y} \in \mathbb{R}^{N\times K}$,  $\bH_A^{(k)} := \bC^{(k-1)} \odot \bB^{(k-1)}$, and  $r_+(\bar{\bA})$ is the regularizer corresponding to the nonnegativity constraint,
\begin{eqnarray*}
  r_+(\bar{\bA}):=\begin{cases}
  0 \text{\hspace{1cm} if \hspace{0.1cm }  $\bar{\bA}\geq \textbf{0}$} \\
  +\infty\hspace{1 cm} \text{ o.w.}
  \end{cases}
\end{eqnarray*}

The ADMM solver then proceeds  by iteratively updating blocks of variables $\bA,\bar{\bA},\bY$ as
\begin{eqnarray}\label{admmA}
\begin{cases}
{\bA}^{(r)}=&\arg\min_{\bA}  \mathcal{L}_A^{(k)}(\bA,\bar{\bA}^{(r-1)},\bY^{(r-1)})\\
\bar{\bA}^{(r)}=& \mathcal{P}_+(\bY^{(r-1)}+\bA^{(r)})\\
%
\bY^{(r)} = &\bY^{(r-1)}- \rho(\bA^{(r)}-\bar{\bA}^{(r)}) \\
r \hspace{0.5cm}=&r+1
\end{cases}
  \end{eqnarray}
until a convergence criterion is met, namely whether  the maximum  number of  iterations is exceeded, i.e.,  $r={I_{\text{max}, \text{ADMM}}}$, or a prescribed $\epsilon$-accuracy is met, i.e., ${\|\bA^{(r)} - \bA^{(r-1)}\|_F}/{\|\bA^{(r-1)}\|_F}<\epsilon$. Operator $\mathcal{P}_+(.)$ in \eqref{admmA} denotes the element-wise projection of the input matrix onto the positive orthant, and its use enables the $\bar{\bA}^{(r)}$ update to be carried at a very low cost.  The Lagrange multiplier is set to $\rho = \|\mathbf{\mathbf{H}}_A\|_F^2/K$- a value that is empirically shown to yield similar performance to that of the optimal value~\cite{keijun}. The final $\bar\bA^{(r)}$ iterate in the ADMM solver will be used to update $\bA^{(k)}$.

\subsubsection{Factor $\bB$ update}
Update of factor $\bB$ can be similarly carried out by solving the subproblem 
\begin{eqnarray}
\label{Bupdate}
 {\bB}^{(k)}   = &\underset{\bB\geq \mathbf{0}}{\text{arg min}} 
\|\bW_2- \bH_B^{(k)}\bB^\top\|_F^2  + \lambda\| \bB \|_F^2
\end{eqnarray}
where $\bW_2:= [\text{vec}(\underline{\bW}_{:,1,:}) , \ldots , \text{vec}(\underline{\bW}_{:,N,:})]$, and $\bH_B^{(k)} := \bC^{(k-1)} \odot \bA^{(k)}$, yielding a similar optimization problem as in \eqref{Aupdate}.
Algorithm \ref{alg:Aupdate} tabulates the explicit update rules for solving \eqref{Aupdate} and similarly \eqref{Bupdate} using a general framework. 

\subsubsection{Factor $\bC$ update}
Update of factor $\bC$ is obtained by fixing $\bA$ and $\bB$ at their most recent values, and solving the subproblem
\begin{eqnarray}\label{Cupdate}
{\bC}^{(k)} = & \arg\min_{\bC}\| {\bW_3} - (\bA^{(k)} \odot \bB^{(k)}) \bC ^\top \|_F^2\\
& \hspace{-1cm}\text{s.t.}  \qquad ,   \bC\geq \mathbf{0}\nonumber   \quad \|\tilde{\bc}_n\|_1=1 \quad \forall n=1,\cdots ,N
\nonumber
\end{eqnarray}
where $\bW_3:= [\text{vec}(\underline{\bW}_{:,:,1}) , \ldots , \text{vec}(\underline{\bW}_{:,:,N})]$. Utilizing an ADMM approach, the augmented Lagrangian is formed as 
\begin{eqnarray}
  \mathcal{L}_C^{(k)}(\bC,\bar{\bC},Y)
  &=  \|\bW_3 - \bH_C^{(k)} \bC^\top \|_F^2\nonumber  +  r_{\text{simp}}(\bar{\bC}) \\   & +({\rho}/2) \|\bY+\bC-\bar{\bC}\|_F^2
\end{eqnarray}
where $\bar{\mathbf{C}},\mathbf{Y} \in \mathbb{R}^{N\times K}$, $\bH_C^{(k)} := (\bA^{(k)} \odot \bB^{(k)})$, and  $r_{\text{simp}}(\bar{\bC})$ is the regularizer corresponding to the simplex constraint on the rows of matrix $\bar{\bC} $ as
\begin{eqnarray*}
  r_{\text{simp}}(\bar{\bC}):=\begin{cases}
  0 \text{\hspace{1cm} if \hspace{0.1cm }  $\bar{\bC} \geq \mathbf{0} , \sum_{k=1}^K{\bar{c}}_{n,k}=1  \; \forall n$} \\
  +\infty\hspace{0.6 cm} \text{ o.w.}
  \end{cases}
\end{eqnarray*}
ADMM solver then proceeds with iterative updates as 
\begin{eqnarray}\begin{cases}
{\bC}^{(r)}=&\arg\min_{\bC}    \mathcal{L}_C^{(k)}(\bC,\bar{\bC}^{(r-1)},\bY^{(r-1)}) \\
\bar{\bC}^{(r)}=& \mathcal{P}_{\text{simp}}(\bY^{(r-1)}+\bC^{(r)})\\
%
\bY^{(r)} = &\bY^{(r-1)} -\rho(\bC^{(r)}-\bar{\bC}^{(r)}) \\
r \hspace{0.5cm}=&r+1
\end{cases}
  \end{eqnarray}
where $\mathcal{P}_{\text{simp}}(.)$ denotes the projection of rows of the input matrix onto the simplex set. This projection has been widely studied and can be efficiently accommodated by the algorithm discussed in \cite{simplex}.
Explicit update steps of the ADMM solver for \eqref{Cupdate}  are tabulated under Algorithm \ref{alg:Cupdate}.

\begin{algorithm}[t]
\caption{{Constrained tensor decomposition via alternating least-squares (ALS)}}
\begin{algorithmic}
\State {Input } $\underline{\bW}, K, I_{\max }, \lambda $
\State {Initialize $\bA, \bB,\bC \in \mathbb{R}^{N \times  K}$ at random and set $k=0$}
\State {Form Matrix reshapes $\bW_1,\bW_2,\bW_3 $ of the tensor as }
\State $\bW_1:= [\text{vec}(\underline{\bW}_{1,:,:}) , \ldots , \text{vec}(\underline{\bW}_{N,:,:})]$
\State $\bW_2:= [\text{vec}(\underline{\bW}_{:,1,:}), \ldots , \text{vec}(\underline{\bW}_{:,N,:})]$
\State $\bW_3:= [\text{vec}(\underline{\bW}_{:,:,1}), \ldots , \text{vec}(\underline{\bW}_{:,:,N})]$
\While {$k<I_{\text{max}}$} or not-convergenced
\State $\mathbf{H}_A^{(k)}=\bC^{(k-1)} \odot \bB^{(k-1)}$
\State { $\mathbf{A}^{(k)}$  $\leftarrow$ Algorithm \ref{alg:Aupdate} with input $\{\bH_A^{(k)} , \bW_1,\bA^{(k-1)}\}$ } 

\State $\mathbf{H}_B^{(k)}=\bC^{(k-1)} \odot \bA^{(k)}$
\State { $\mathbf{B}^{(k)}$  $\leftarrow$ Algorithm \ref{alg:Aupdate} with input $\{\bH_B^{(k)} , \bW_2,\bB^{(k-1)}\}$ } 
\State $\mathbf{H}_C^{(k)}=\bB^{(k)} \odot \bA^{(k)}$
\State { $\mathbf{C}^{(k)}$  $\leftarrow$ Algorithm \ref{alg:Cupdate} with input $\{\bH_C^{(k)} , \bW_3,\bC^{(k-1)}\}$ } 
\State ${k \leftarrow k+1}$
\EndWhile
\State \textbf{Retrun} ${\bA}^{(k)},{\bB}^{(k)},{\bC}^{(k)}$ 

\end{algorithmic}
\label{alg:ALS}
\end{algorithm}

\begin{algorithm}[t]
\caption{{ADMM solver for $1^{\text{st}}$ and $2^{\text{nd}}$ mode subproblems}}
\begin{algorithmic}
\State {Input }  ${\bH},\bW, \bZ_{\text{init}}  $

\State {\bf{Goal is to solve}} \begin{eqnarray}
\nonumber
{\bZ}^*   = &\underset{\bZ \geq \mathbf{0}}{\text{arg min}} 
\text{Tr}\Big{\{}\bZ (\bH^\top \bH  + \lambda\mathbf{I}_{K\times K })\bZ^\top  -2\bW^\top \bH \bZ^\top\Big{\}}
\end{eqnarray}

\State {Set $  \rho = \dfrac{\|\bZ_{\text{init}}\|_F^2}{K} , \bZ^{(0)}= \bZ_{\text{init}} , \Bar{\bZ}^{(0)} = \mathbf{0}_{{N \times K}}, \bY^{(0)}  = \mathbf{0}_{{N \times K}}, r=0 $}
\While {$r<I_{\text{max,ADMM}}$}
\State{\begin{equation*}
	\begin{aligned}
	&{\bZ}^{(r)}  = (\bH^\top \bH   + (\lambda + \rho/2 ) \mathbf{I}_{K\times K })^{-1} \\  & \hspace{1 cm}\times \Big(\bW^\top \bH+\dfrac{\rho}{2}(\bY^{(r-1)} -\bar{\bZ}^{(r-1)}) \Big) 
	\end{aligned}
	\end{equation*}}
\State {$\hspace{0.5cm}\bar{\bZ}^{(r)}= \mathcal{P}_+\Big(\bZ^{(r)}+\mathbf{Y}^{(r-1)}\Big)$}
\State $ \hspace{0.5cm}\bY^{(r)} = \bY^{(r-1)} - \rho(\bZ^{(r)}-\bar{\bZ}^{(r)}) \nonumber$
\State $\hspace{0.5cm} r \hspace{0.5cm}=r+1$
\EndWhile
\State \textbf{Retrun} $\bar{\bZ}^{(r)}$ 
\end{algorithmic}
\label{alg:Aupdate}
\end{algorithm}

\begin{algorithm}[t]
\caption{{ADMM solver for $3^{\text{rd}}$ mode subproblem}}
\begin{algorithmic}
\State {\bf Input }  ${\bH},\bW, \bZ_{\text{init}}  $

\State {\bf{Goal is to solve}} \begin{eqnarray}
\nonumber
{\bZ}^*   = &\underset{\bZ\geq \mathbf{0}\|\tilde{\mathbf{z}}_n\|_1 =1 \; \forall n=1,\cdots,N }{\text{arg min}} 
\text{Tr}\Big{\{}\bZ \bH^\top \bH \bZ^\top -2\bW^\top \bH \bZ^\top\Big{\}}
\end{eqnarray}
\State {Set $  \rho = \dfrac{\|\bZ_{\text{init}}\|_F^2}{K} , \bZ^{(0)}= \bZ_{\text{init}} , \Bar{\bZ}^{(0)} = \mathbf{0}_{{N \times K}}, \bY^{(0)}  = \mathbf{0}_{{N \times K}}, r=0 $}
\While {$r<I_{\text{max,ADMM}}$}
\State \begin{equation}
\begin{aligned}
&{\bZ}^{(r)}= (\bH^\top \bH+\rho/2 \, \mathbf{I}_{N\times N})^{-1} \nonumber  \\   & \hspace{1cm} \times\Big(\bW^\top \bH+\dfrac{\rho}{2}(\bY^{(r-1)}-\bar{\bZ}^{(r-1)}) \Big)
\end{aligned}
\end{equation}
\State {\hspace{0.5cm} $\bar{\bZ}^{(r)}= \mathcal{P}_{\text{simp}}\Big(\bZ^{(r)}+\bY^{(r-1)}\Big)$}
\State \hspace{0.5cm} $\bY^{(r)} = \bY^{(r-1)} -\rho(\bZ^{(r)}-\bar{\bZ}^{(r)}) \nonumber$
\State \hspace{0.5cm} $r \hspace{0.5cm}=r+1$
\EndWhile
\State \textbf{Retrun} $\bar{\bZ}^{(r)}$ 
\end{algorithmic}
\label{alg:Cupdate}
\end{algorithm}

{\bf{Proposition}}
If the sequence generated by Alg. \ref{alg:ALS} is bounded, then the sequence $\{\bA^{(k)}, \bB^{(k)}, \bC^{(k)}\}$ converges to a stationary point of \eqref{parafac2}.

{\bf{Proof:}} The convergence follows from   \cite[Theorem 1]{keijun}.

\section{Community assignment and  evaluation }
\label{sec:metrics}
Once the proposed solver returns the solution of \eqref{parafac2}, the rows of factor $\hat{\bC}$ provide a ``soft'' or ``fuzzy'' community membership for the nodes in the network. In the special case of networks with non-overlapping communities, using the $n$-th row of matrix $\hat{\bC}$,  node $n$ will be assigned to the community $k^*$ where $k^* = \arg\underset{{k=1,\cdots,K}}\max {\hat{c}}_{nk}$.

In order to provide a ``crisp'' community association in networks {\it with} overlapping communities, where a node can be associated with more than one community, the entries $\hat{c}_{nk}$ are compared with a threshold $\tau$ and node $n$ is associated with community $k$ if $\hat{c}_{nk} > \tau$. Thus, crisp community membership matrix $\boldsymbol{\Gamma} \in \mathbb{R}^{N \times K}$ is obtained as 
\begin{equation}
[\boldsymbol{\Gamma}]_{nk} :=
\begin{cases}
1 \text{\qquad  if $ \; \hat{c}_{nk}>\tau$} \\
0 \qquad \text{o.w.}
\end{cases} \qquad \forall {n,k}
\end{equation}  

There are a number of metrics available for evaluating the quality of a detected \emph{cover}, that is, a set of communities, in networks with underlying community structure. Depending on whether the ground-truth communities are available or not, two categories of metrics are considered. 
\subsection{Networks with ground-truth communities}

Normalized mutual information and F1-score are the most commonly-used metrics for performance evaluation over networks with ground truth communities. Let cover $\hat{\mathcal{S}}:= \{\hat{\mathcal{C}}_1, 
\ldots, \hat{\mathcal{C}}_{|\hat{\mathcal{S}}|}\}$ denote the set of detected communities, where $\mathcal{C}_i$  is the set of  nodes associated with community $i$ for $i=1,2,\ldots,|\hat{\mathcal{S}}|$, and let the ground truth communities be denoted by $\mathcal{S}^* :=\{{\mathcal{C}}^*_1, 
\ldots, {\mathcal{C}}^*_{|\mathcal{S}^*|}\}$. 

\noindent{\bf Normalized mutual information (NMI)~\cite{Fortu}:} 
 NMI is an information-theoretic metric defined as  (cf.~\cite{Fortu})
\[
\text{NMI}(\mathcal{S}^*,\hat{\mathcal{S}}) := \dfrac
{2 \text{I}(\mathcal{S}^*,\hat{\mathcal{S}}) }{\text{H}(\mathcal{S}^*)+\text{H}(\hat{\mathcal{S}}) }
\] 
where $\text{H}(\hat{\mathcal{S}})$ denotes the entropy of set $\hat{\mathcal{S}}$ defined as   \[\text{H}(\hat{\mathcal{S}}):= -\sum_{i=1  }^{|\hat{\mathcal{S}}|} p(\hat{\mathcal{C}}_i)\log p(\hat{\mathcal{C}}_i)= -\sum_{i=1  }^{|\hat{\mathcal{S}}|}   \dfrac{|\hat{\mathcal{C}}_i|}{N}\log \dfrac{|\hat{\mathcal{C}}_i|}{N}\]
and similarly for  $\text{H}({\mathcal{S}}^*)$. Furthermore,   $\text{I}(\mathcal{S}^*,\hat{\mathcal{S}}) $     denotes the mutual information between the detected and ground-truth communities,     and is defined as 
\begin{eqnarray}
\text{I}(\mathcal{S}^*,\hat{\mathcal{S}}) := \sum_{i=1}^{|\mathcal{S}^*|} \sum_{j=1}^{|\hat{\mathcal{S}}|} p(\mathcal{C}^*_i\cap \hat{\mathcal{C}}^*_j)
\log\dfrac{p(\mathcal{C}^*_i\cap \hat{\mathcal{C}}_j)}{p(\mathcal{C}^*_i)p( \hat{\mathcal{C}}_j)} \\
= \sum_{i=1}^{|\mathcal{S}^*|} \sum_{j=1}^{|\hat{\mathcal{S}}|} \dfrac{|\mathcal{C}^*_i\cap \hat{\mathcal{C}}_j|}{N}
\log\dfrac{N|\mathcal{C}^*_i\cap \hat{\mathcal{C}}_j|}{|\mathcal{C}^*_i| | \hat{\mathcal{C}}_j|} 
\end{eqnarray}
Intuitively,  mutual information $\text{I}(\mathcal{S}^*,\hat{\mathcal{S}}) $ reflects a measure of similarity between the two community sets, while entropy
$\text{H}(\hat{\mathcal{S}})$ $(\text{H}({\mathcal{S}}^*))$ denotes the level of uncertainty in community affiliation of a random node in cover $\hat{\mathcal{S}}$ (resp. $\mathcal{S}^*$).  Thus, high values of NMI, namely its maximum at 1,  reflect \textit{predictability} of $\hat{\mathcal{S}}$ based on ${\mathcal{S}^*}$ which readily translates into correct community identification in the detected cover $\hat{\mathcal{S}}$, whereas low values of NMI, namely its minimum at 0, reflects poor discovery of the true underlying communities.
This measure has been generalized for overlapping communities in \cite{GNMI}, and will be utilized for performance assessment in such networks.

\noindent{\bf  Average F1-score} \cite{bigclam}: F1-score is a measure of binary classification accuracy, 
specifically, the harmonic mean of \textit{precision} and \textit{recall}, taking its highest value at 1 and lowest value at 0. 
To obtain the average F1-score for $\hat{\mathcal{S}}$, one needs to find which detected community $\hat{\mathcal{C}}_i  \in \hat{\mathcal{S}} $ 
corresponds to a given true community ${\mathcal{C}}_j^* \in {\mathcal{S}^*}$, i.e., maximizes the corresponding F1-score. 
  The average F1-score is then given by 
\[ \bar{F1}:= \dfrac{1}{|\mathcal{S}^*|}  \sum_{i=1}^{|\mathcal{S}^*|} F1({\mathcal{C}^*_i}, \hat{\mathcal{C}}_{I(i)}) 
\]
where 
\[I(i) = \arg\max_j F1(\mathcal{C}^*_i,\hat{\mathcal{C}}_j) \]
in which $F1(\mathcal{C}^*_i,\hat{\mathcal{C}}_j):= \dfrac{2 \, |\mathcal{C}^*_i\cap\hat{\mathcal{C}}_j|}{|\mathcal{C}^*_i|+|\hat{\mathcal{C}}_j|}.$

\subsection{Networks without ground-truth communities}

A general metric for evaluating the ``quality'' of detected communities, regardless of whether the ground-truth memberships are available or not, is to measure  \textit{conductance}~\cite{leskovec2009conductance}.

{\bf Conductance}:
Conductance of a detected community $\hat{\mathcal{C}}_k$ in graph $\mathcal{G}$ is defined as
\begin{equation*}
\phi(\hat{\mathcal{C}}_k) := \dfrac{\sum_{i \in \hat{\mathcal{C}}_k,j\notin \hat{\mathcal{C}}_k}\bW_{ij}}{\min\{\text{vol}({\hat{\mathcal{C}}_k}) ,\text{vol}(\mathcal{V}\setminus{{\hat{\mathcal{C}}}}_k)\}}
\end{equation*}
where 
\begin{equation*}
\text{vol}({{\hat{\mathcal{C}}_k}}):= {\sum_{i \in \hat{\mathcal{C}}_k,\forall j }\bW_{ij}}
\end{equation*}
 and $(\mathcal{V}\setminus{{\hat{\mathcal{C}}}}_k)$ is the complement of $\hat{\mathcal{C}}_k$.
According to this measure, \textit{high-quality} communities yield small conductance scores as they exhibit dense connections among the nodes within the community and sparse connections with the rest.

Furthermore, the weighted-average $\bar{\phi}(\hat{\mathcal{S}})$ is defined as  the average conductance of the detected communities weighted by their ({normalized}) community size, that is, 
\begin{equation}\label{average_conductance}
\bar{\phi}(\hat{\mathcal{S}}) := \sum_{k=1}^{|\hat{\mathcal{S}}|}\dfrac{|\hat{\mathcal{C}}_k|}{N} \phi(\hat{\mathcal{C}}_k).
\end{equation}

\section{Community detection on time-varying graphs }\label{sec:timevarying}
In this section, we extend the introduced overlapping community identification approach over networks for which the connectivity evolves over time. For instance, consider the emergence of a new sports club  giving rise to a new community of individuals  whose newly-formed interactions in the  club will  be reflected in their connections over the social media, or, the  network of brain regions where the activation/deactivation of different regions during a certain task can be captured by a time-varying graph.  The goal  is to utilize the proposed EgoTen approach for identification of  dynamic communities, as well as  the corresponding time-varying association of nodes.

To this end, consider  graph $\mathcal{G}_t:=(\mathcal{V},\mathcal{E}_t,\mathbf{W}_t)$, where the subscript $t$ denotes time index $t=1,...,T$. The set of nodes $\mathcal{V}$ is  assumed  fixed across time, while the edgeset $\mathcal{E}_t$ as well as the corresponding adjacency matrix $\bW_t$ are allowed to vary.  The introduced EgoTen community identification algorithm can be readily applied to   time-varying graphs as follows. 

For any slot $t$, the procedure of egonet tensor construction can be carried out, giving rise to a $3$-dimensional egonet-tensor denoted by $\underline{\bW}_t$. Subsequently, the overall $4$-dimensional egonet-tensor is  constructed by stacking $\underline{\bW}_t \; \forall t$ along the fourth dimension of $\underline{\bW} \in \mathbb{R}^{N \times N\times N \times T}$; that is, according to the tensor parlance we have $\underline{\bW}_{:,:,:,t} = \underline{\bW}_t$ for $t=1,\ldots,T$. 
Having formed the overall egonet-tensor $\underline{\bW}$, dynamic community detection is now cast as
\begin{eqnarray}
\label{parafac4d}
\nonumber
 \{ \widehat{\bA}, \widehat{\bB}, \widehat{\bC} ,\widehat{\bD} \}   = &\hspace{-0.9cm} \underset{\bA,\bB,\bC,\bD}{\text{arg min}} 
\Big\{\|\underline{\bW} - \sum_{k=1}^K \ba_k \circ \bb_k\circ \bc_k \circ \mathbf{d}_k \|_F^2 \hspace{-1cm}\\ 
&+ \lambda(\|\bA\|_F^2+\|\bB\|_F^2)\Big\}\\
& \text{s.t.}  \qquad   \bA\geq \mathbf{0},\bB\geq \mathbf{0},\bC \geq \mathbf{0},\bD\geq \mathbf{0}\nonumber  \\
 &\|\tilde{\bc}_n\|_1=1  \qquad  \forall n=1,2,...,N \nonumber \\ \nonumber
 &  \|\tilde{\mathbf{d}}_n\|_1=1  \qquad  \forall n=1,2,...,N 
\end{eqnarray}
where $\bA,\bB,\bC \in \mathbb{R}^{N \times K}$, and $\bD := [\tilde{\mathbf{d}}_1^\top,\ldots,\tilde{\mathbf{d}}_T^\top]^\top \in \mathbb{R}^{T \times K}$. The LS cost in \eqref{parafac4d}  is the generalization of that for the $3$-dimensional tensor decomposition to a higher dimension, while nonnegativity and simplex constrains are similarly carried over for  $\bD$. To clarify the simplex constraints on the rows of  $\bD$, consider the decomposition 
\begin{equation}
\underline{\bW}_t = \sum_{k=1}^K {d}_{tk} \Big(\ba_k \circ \bb_k \circ \bc_k\Big).
\end{equation}

For a fixed slot $t$, the rank-one tensors of $(\ba_k \circ \bb_k \circ \bc_k)$ can be perceived as  building blocks of the $3$-dimensional egonet-tensors, and the  ``degree of presence'' of community $k$ at time $t$ is captured by $d_{kt}$,  while the constraint $\|\tilde{\mathbf{d}}_t\|_1=1 \; \forall t$ resolves the scalar ambiguity by normalization. Indeed, stationary graphs with $T=1$ can be subsumed by this model, for which the additional constraints on $\bD$ reduce to the trivial solution for the fourth factor as  $\bD=\mathbf{1}_{1 \times K}$. 
Focusing on the $n$-th slab of $\underline{\bW}_t$, the egonet  adjacency of node $n$ at time $t$ is decomposed as 
\begin{equation}
{\bW}_t^{(n)} = \sum_{k=1}^K {d}_{tk} c_{nk} \Big(\ba_k \circ \bb_k\Big)
\end{equation}
where the product $d_{tk} c_{nk}$ provides the association of node $n$ to community $k$ at time $t$.

Similar to \eqref{parafac2}, the decomposition in \eqref{parafac4d} is solved by alternating least-squares for updating the factors. The overall solver is provided in Alg. 4, within which we have utilized  Alg. 2 and Alg. 3 for handling the emerging subproblems.

\begin{algorithm}[t]
\caption{{Constrained ALS for time-varying graphs}}
\begin{algorithmic}
\State {Input } $\underline{\bW}, K, I_{\max }, \lambda $
\State  {Initialize $\bA, \bB,\bC \in \mathbb{R}^{N \times  K}$ and $\bD \in \mathbb{R}^{T \times  K}$   and set $k=0$}
\State {Form matrix reshapes $\bW_1,\bW_2,\bW_3 , \bW_4$ of the tensor as }
\State $\bW_1:= [\text{vec}(\underline{\bW}_{1,:,:,:}) , \ldots , \text{vec}(\underline{\bW}_{N,:,:,:})]$
\State $\bW_2:= [\text{vec}(\underline{\bW}_{:,1,:,:}) , \ldots , \text{vec}(\underline{\bW}_{:,N,:,:})]$
\State $\bW_3:= [\text{vec}(\underline{\bW}_{:,:,1,:}) , \ldots , \text{vec}(\underline{\bW}_{:,:,N,:})]$
\State $\bW_4:= [\text{vec}(\underline{\bW}_{:,:,:,1}) , \ldots , \text{vec}(\underline{\bW}_{:,:,:,T})]$
\While {$k<I_{\text{max}}$} or not-convergenced
\State $\mathbf{H}_A^{(k)}=\bD^{(k-1)} \odot \bC^{(k-1)} \odot \bB^{(k-1)}$
\State { $\mathbf{A}^{(k)}$  $\leftarrow$ Algorithm \ref{alg:Aupdate} with input $\{\bH_A^{(k)} , \bW_1,\bA^{(k-1)}\}$ } 

\State $\mathbf{H}_B^{(k)}=\bD^{(k-1)} \odot \bC^{(k-1)} \odot \bA^{(k)}$
\State { $\mathbf{B}^{(k)}$  $\leftarrow$ Algorithm \ref{alg:Aupdate} with input $\{\bH_B^{(k)} , \bW_2,\bB^{(k-1)}\}$ } 
\State $\mathbf{H}_C^{(k)}=\bD^{(k-1)} \odot \bB^{(k)} \odot \bA^{(k)}$
\State { $\mathbf{C}^{(k)}$  $\leftarrow$ Algorithm \ref{alg:Cupdate} with input $\{\bH_C^{(k)} , \bW_3,\bC^{(k-1)}\}$ } 
\State $\mathbf{H}_D^{(k)}=\bC^{(k)} \odot \bB^{(k)} \odot \bA^{(k)}$
\State { $\mathbf{D}^{(k)}$  $\leftarrow$ Algorithm \ref{alg:Cupdate} with input $\{\bH_D^{(k)} , \bW_4,\bD^{(k-1)}\}$ } 
\State ${k \leftarrow k+1}$
\EndWhile
\State \textbf{Retrun} ${\bA}^{(k)},{\bB}^{(k)},{\bC}^{(k)},{\bD}^{(k)}$ 

\end{algorithmic}
\label{alg:ALS}
\end{algorithm}

\section{Simulated tests}
\label{sec:tests}
In this section, the performance of the proposed EgoTen community detection algorithm is assessed via benchmark synthetic networks, as well as real-world datasets. Experiments over Lancicchinetti-Fortunatoand-Radicci (LFR) synthetic benchmarks enables us to simulate networks with different levels of community mixing, as well as  number of overlapping nodes. This proves helpful in highlighting the enhanced capacity of  community detection achieved by exploiting the ``higher-order'' properties of vertices captured in the proposed egonet-tensor.

\subsection{Benchmark networks}
\begin{figure*}
	\vspace{-2cm}
	\begin{minipage}[b]{0.5\linewidth}
		\centering
		\centerline{\includegraphics[width=0.9\textwidth]{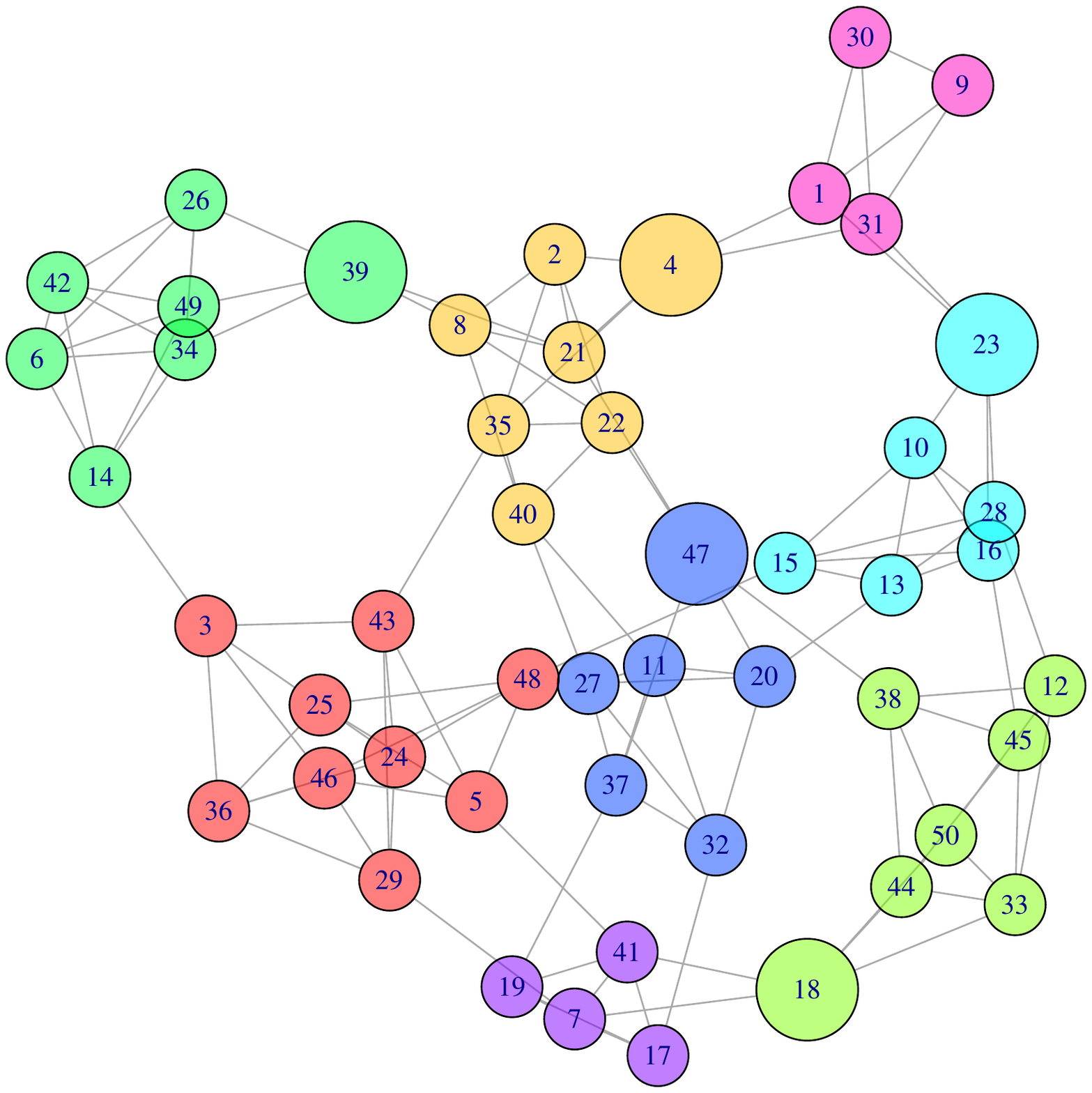}\vspace{-2cm}}
		\centerline{(a) }\medskip
	\end{minipage}
	\begin{minipage}[b]{.5\linewidth}
		\centering
		\centerline{\includegraphics[width=0.9\textwidth]{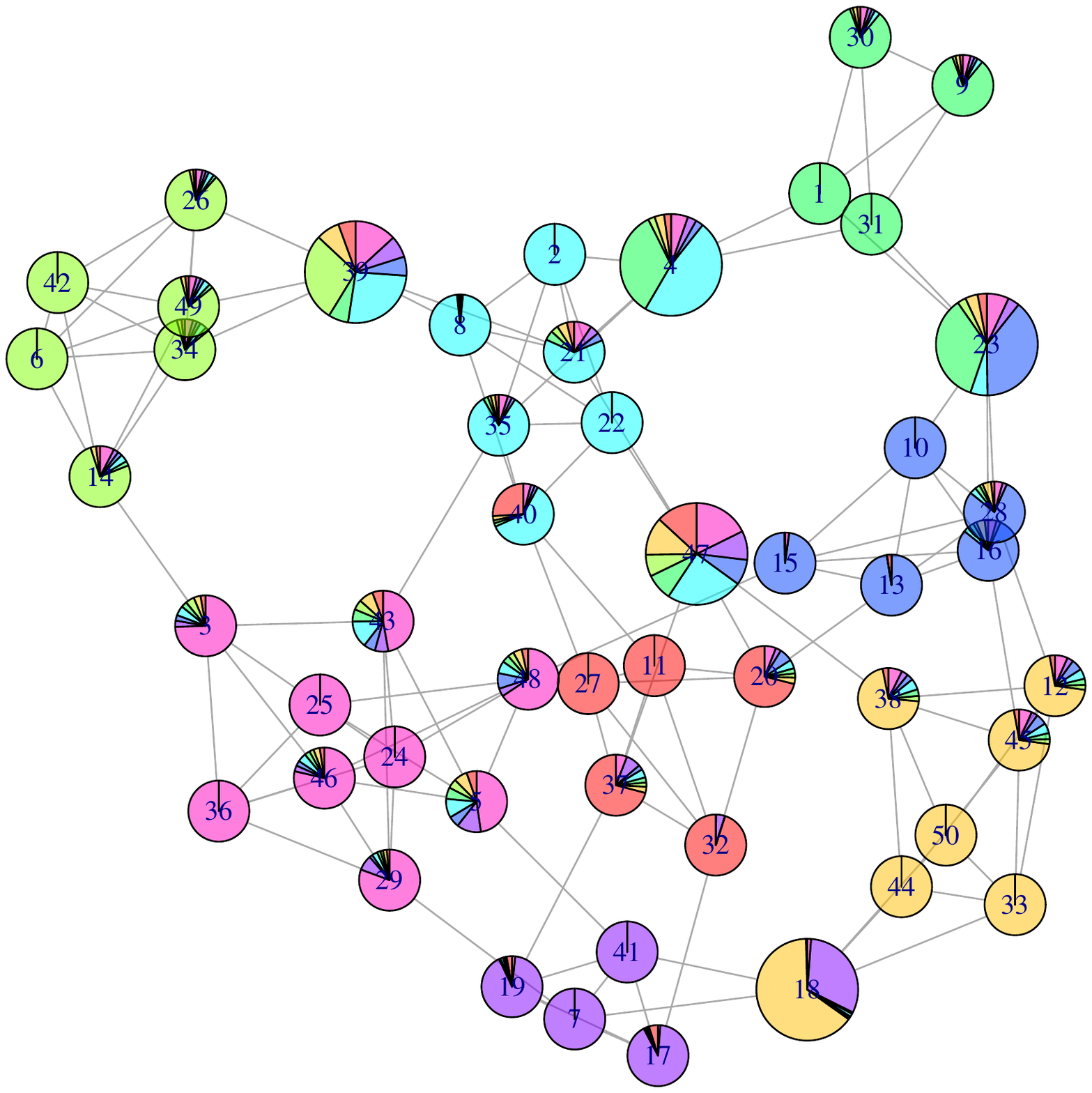}\vspace{-2cm}}
		\centerline{(b) }\medskip
	\end{minipage}

	\caption{Visualization of an LFR network with $N=50$,  $\mu=0.2$, and five ``shared'' nodes $\{4,18,23,39,47\}$, represented by a larger radius, with (a) hard community association via Infomap; and (b) soft association via EgoTen. Pie-charts depict association ratios. }
\end{figure*}
 LFR  benchmark networks provide synthetic graphs with ground-truth communities, in which certain properties of real-world networks, namely power-law distribution for nodal degrees as well as community sizes are preserved. 
LFR networks are  configured to have a total number of $N$ nodes (vertices), nodal average degree $\bar{d}$, exponent of degree distribution $\gamma_1$, and exponent of community-size distribution $\gamma_2$. Furthermore,  \emph{mixing parameter} $\mu$ controls the community cohesion, where larger $\mu$ induces more connections among nodes in different communities, thus generating less cohesive communities.  Moreover, parameters $\{o_n,o_m\}$ set the  number of overlapping nodes, that is, nodes belonging to more than one community, and the number of  communities with which these nodes are associated, respectively. 
We have compared the performance of the proposed EgoTen with (similarly-constrained) nonnegative matrix factorization  (NMF) schemes over the adjacency matrix, as well as other state-of-the-art community detection schemes.

\subsubsection{Matrix versus tensor factorization}
The experiments here focus on comparison between EgoTen and its matrix  counterpart. In particular, we will demonstrate how the higher-order connectivity patterns as well as structured redundancy offered through EgoTen can increase the robustness of community detection in the case of overlapping, and highly-mixed communities. 

To this end, the corresponding NMF approach aims at factorizing matrix $\bW$ by solving
\begin{equation}\label{NMF}
\begin{aligned}
&\min_{\bU,\bV} {\| \bW - \bU \bV^\top\|_F^2} \\
& {\text{s.t.}} \qquad \| \mathbf{u}_n \|_1=1 \, \forall n=1,\ldots,N,\; \bU\geq \textbf{0} \;, \bV \geq \textbf{0}
\end{aligned}
\end{equation}
where $\bU,\bV \in \mathbb{R}^{N \times K}$, and $\bU^\top := [\bu_1,\bu_2,...,\bu_N]$. Thus, similar to the nonnegative tensor decomposition in \eqref{parafac2}, the $n$-th row of matrix $\bU$ contains community association coefficients of node $n$, and is subject to a simplex constraint. This minimization is solved via the AO-ADMM toolbox in \cite{keijun}. Hard community assignments resulting from the factor $\bU$ can be achieved similar to the procedure over the $\hat{\bC}$ factor in EgoTen, as discussed in Section IV.

In our experiments,  we have generated LFR networks with $N=1,000$,  $\gamma_1=2$,  $\gamma_2=1$, average degree $\bar{d}=100$, and $o_m=\{2,3,5\}$. To demonstrate the robustness of Egot-Ten, Fig. 5  shows the performance of EgoTen in comparison with constrained NMF over the adjacency matrix  versus different values of $\mu$, in terms of the NMI, and the F1-score.  The experiments have been carried out for  $o_n=200$ ($20\%$) number of overlapping nodes, upperbound on the number of communities is set as $K=3|\mathcal{C}|$, and $\tau$ is chosen so that the highest NMI is achieved for both methods. 
Furthermore, Fig. 6 depicts the performance of NMF and EgoTen across different levels of overlap, controlled through the number of overlapping nodes $o_n$, while fixing $\mu=0.2$. In addition, since a common practical concern is the lack of knowledge over the number of communities, Fig. 7 studies the robustness of NMF and EgoTen to this factor, by setting $K=m |\mathcal{C}|$, and varying  $m \in [1.2 , 5]$.
As the plots in Figures 5-7 corroborate,  capturing the  higher-order connectivity patterns of the vertices  and the structured decomposition of reinforced egonet-tensor improve robustness of nonnegative factorization methods against community coherence, presence of overlapping nodes, as well as rough estimates of the number of communities.

\begin{figure}
	
	\begin{minipage}[b]{1\linewidth}
		\centering

		\centerline{\includegraphics[width=1\textwidth]{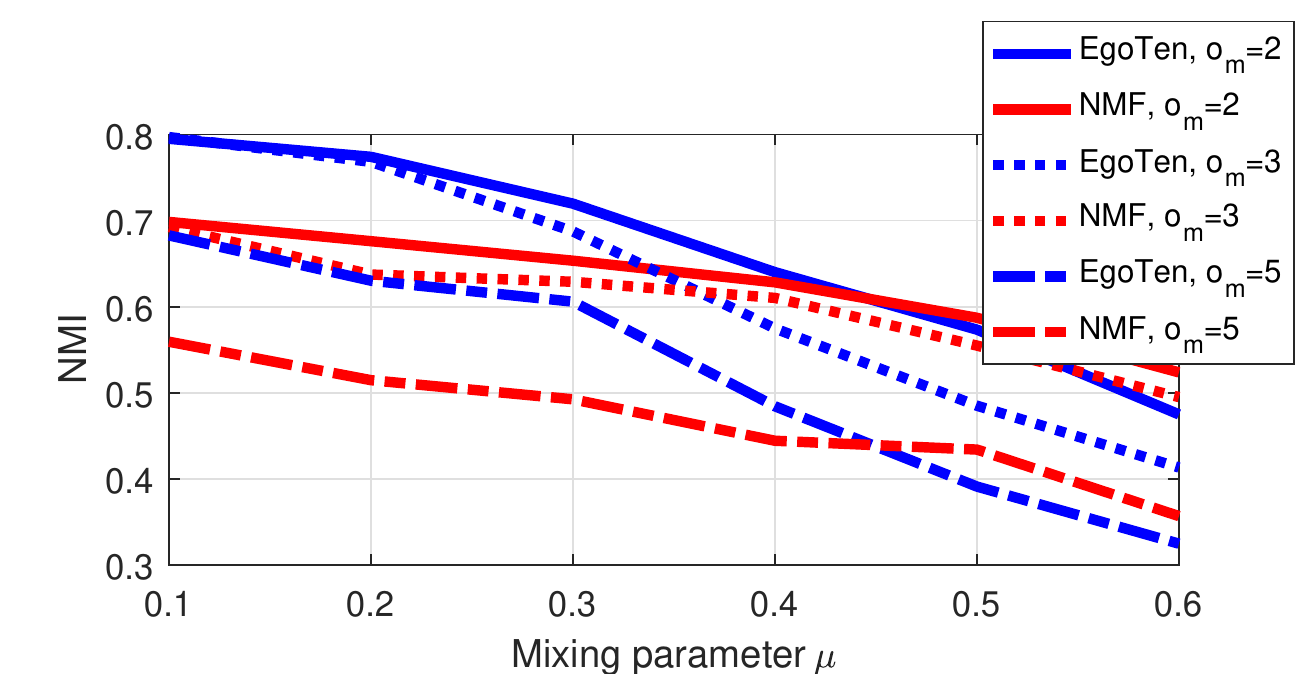}}

		\centerline{(a)   }\medskip
	\end{minipage}

	\begin{minipage}[b]{1\linewidth}
		\centering
		
		\centerline{\includegraphics[width=1\textwidth]{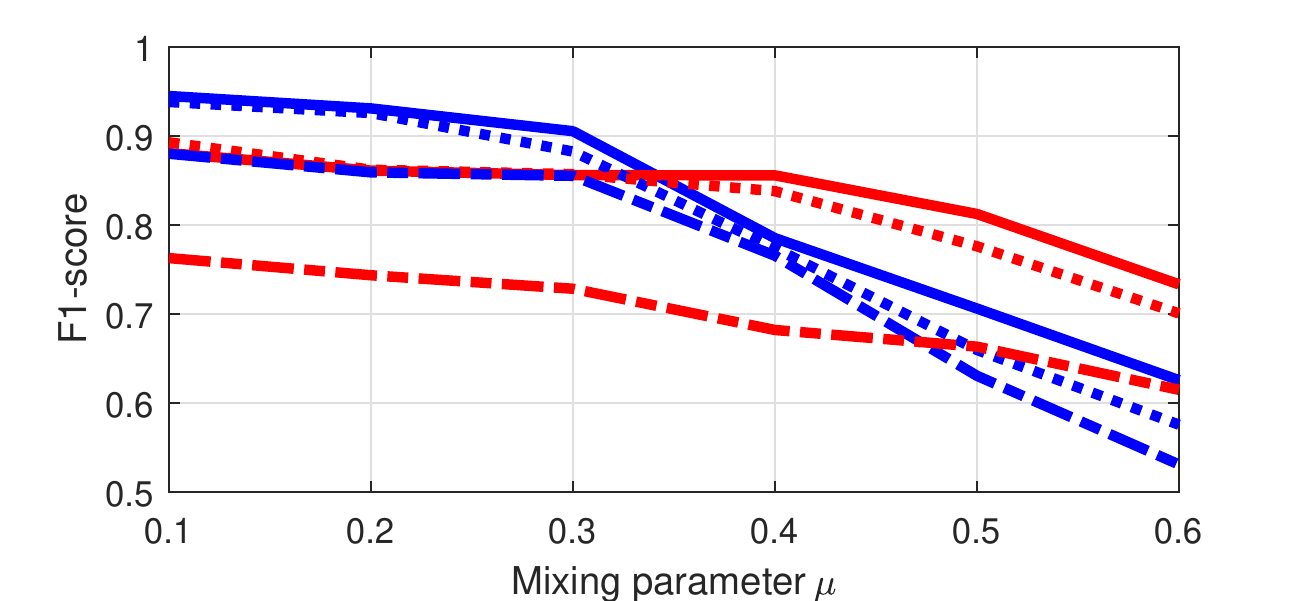}}
		
		\centerline{(b) }\medskip
	\end{minipage}
	\hfill
	\caption{Performance  of constrained NMF and EgoTen in terms of (a) NMI; and, (b) average F1-score, versus  $\mu$ for LFR networks of $N=1,000$, with $o_n=200$.}
	\label{fig:res}
\end{figure}

\begin{figure}
	\vspace{0.0cm}
	\begin{minipage}[t]{1\linewidth}
		\centering
		\centerline{\includegraphics[width=1\textwidth]{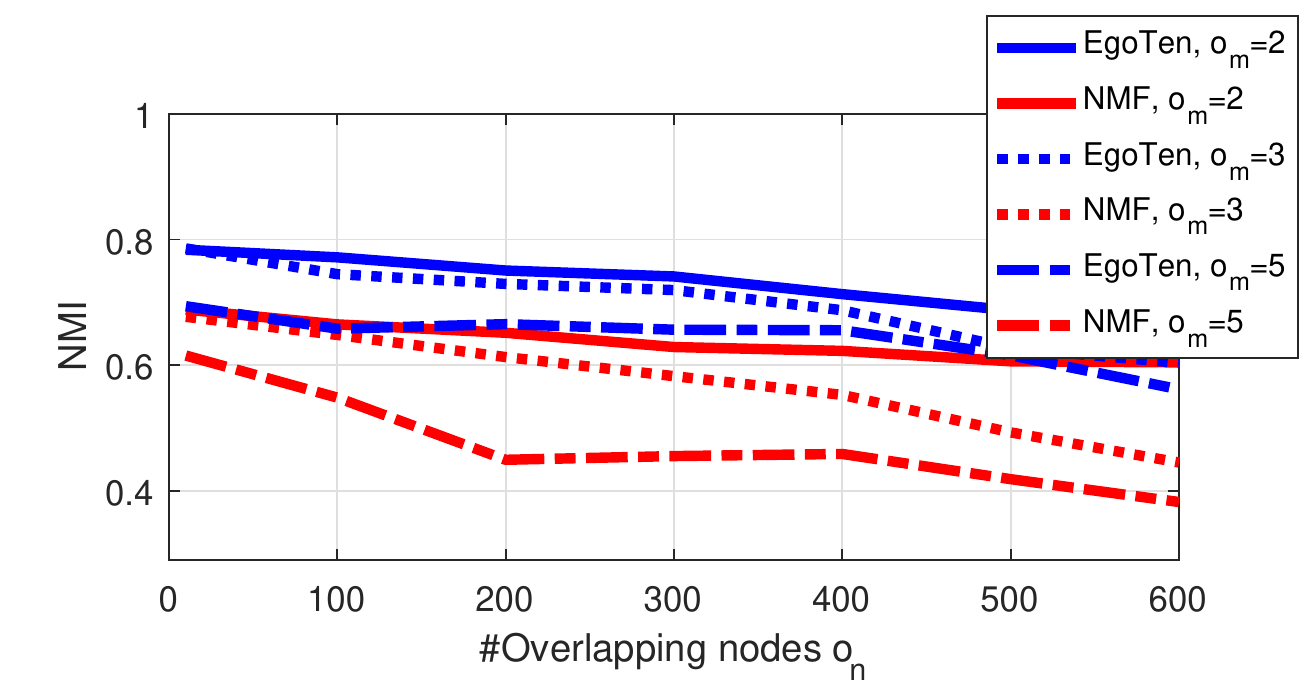}}
		\centerline{(a)   }\medskip

	\end{minipage}

	\begin{minipage}[b]{1\linewidth}
		\centering

		\centerline{\includegraphics[width=1\textwidth]{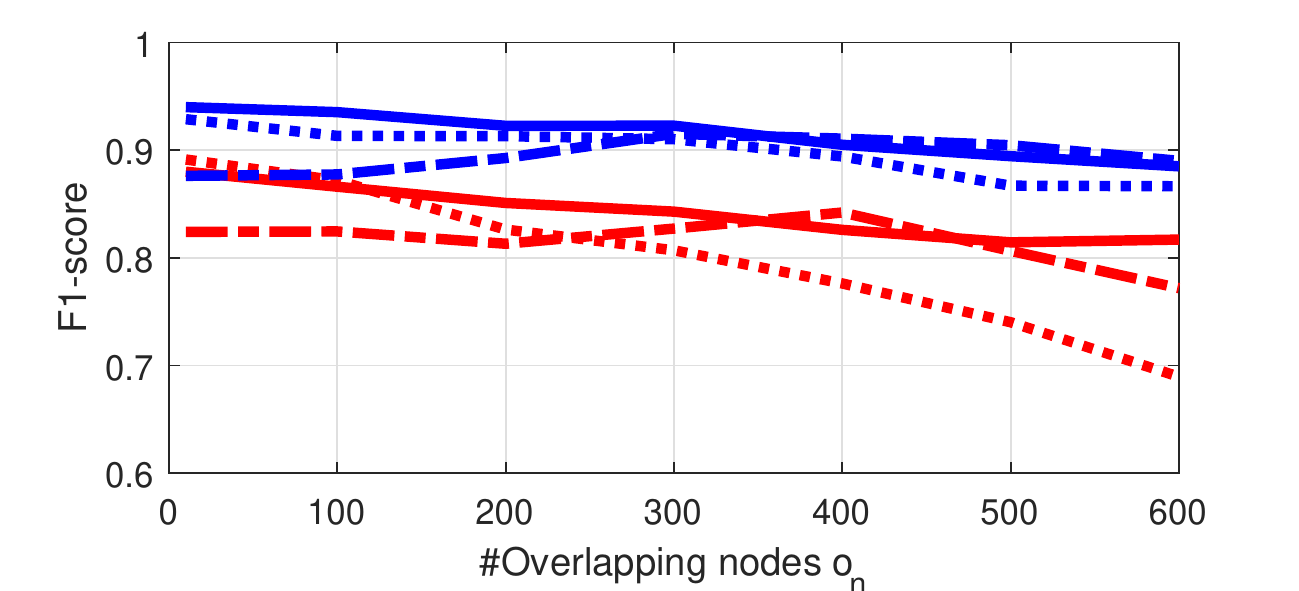}}
		\centerline{(b) }\medskip
	\end{minipage}
	\hfill
	\caption{Performance of constrained NMF and EgoTen in terms of (a) NMI; and, (b) average F1-score,  versus $o_n$ for LFR networks with $N=1,000$, and  $\mu=0.2$.}
	\label{fig:res}
\end{figure}

\begin{figure}

	\begin{minipage}[b]{1\linewidth}
		\centering
		\centerline{\includegraphics[width=1\textwidth]{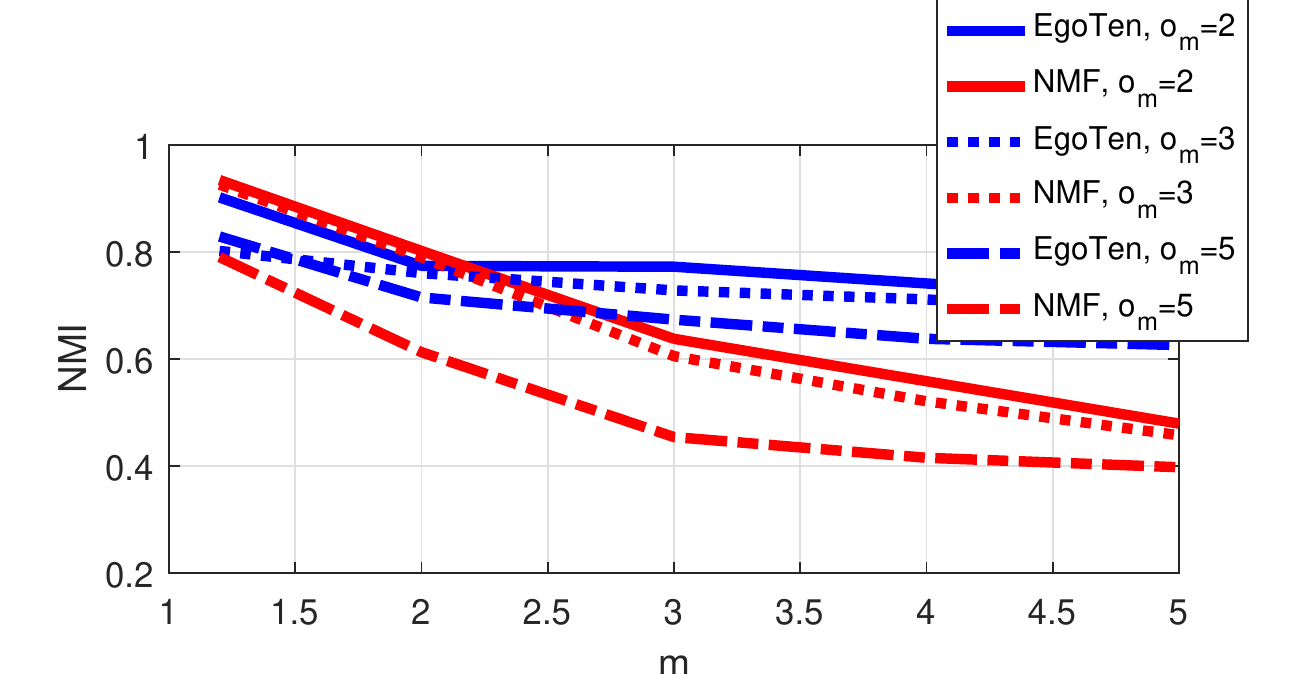}}
		
		\centerline{(a)   \medskip}
			\vspace{-0.2cm}
	\end{minipage}

	\begin{minipage}[b]{1\linewidth}
		\centering
		\vspace{-0.0cm}		
		\centerline{\includegraphics[width=1\textwidth]{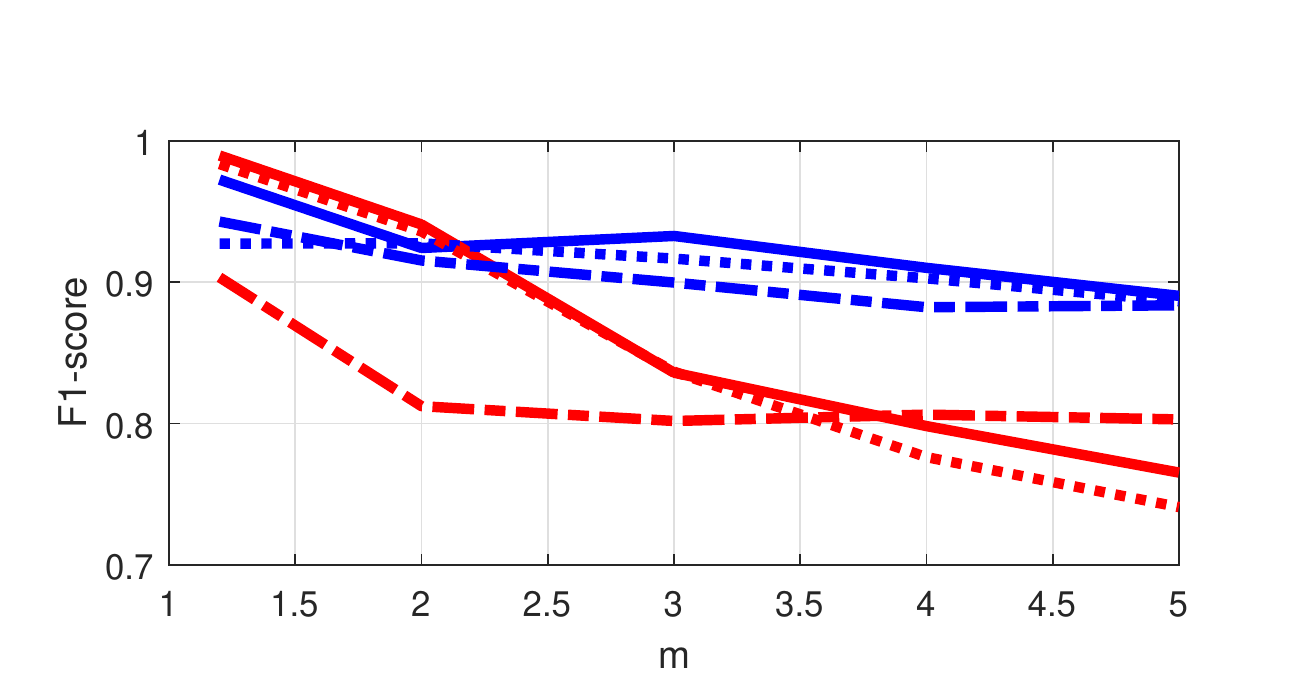}}
		\centerline{(b) }\medskip
	\end{minipage}
	\hfill
	\caption{Performance of constrained NMF  and EgoTen in terms of (a) NMI; and, (b) average F1-score, versus upperbound on community number $K=m|\mathcal{C}|$ parametrized by $m$ for LFR networks with $N=1,000$, $o_n=300$ and $\mu=0.2$.}
	\label{fig:res}
\end{figure}

\subsubsection{EgoTen vs. state-of-the-art methods}

In this subsection, we compare the performance of EgoTen with  state-of-the-art competitors, namely Oslom \cite{oslom}, Bigclam\cite{bigclam}, Infomap\cite{infomap}, and Louvain\cite{louvain}. Similar to the previous subsection, 
we have generated LFR networks with $N=1,000$, $\gamma_1=2$, $\gamma_2=1$ and $o_m=\{2,3,5\}$. Number of overlapping nodes $o_n$ as well as mixing parameter $\mu$ are varied in the range $[10,600]$  and $[0.1,0.7]$, respectively. 
Threshold parameter  for assigning hard community memberships of EgoTen is chosen as\footnote{Since we have imposed a simplex constraint over the $K$ association indices for any given node, $\tau=1/K$ could be interpreted as having an association index higher than an equal association with all detected communities. Other selection schemes for $\tau$, for instance setting to the value providing the community cover with the smallest average conductance, are also viable.} $\tau=1/K$.

\begin{figure}
	\begin{minipage}{1\linewidth}
		\centering
		\vspace{0.0cm}
		\centerline{\includegraphics[width=01\textwidth]{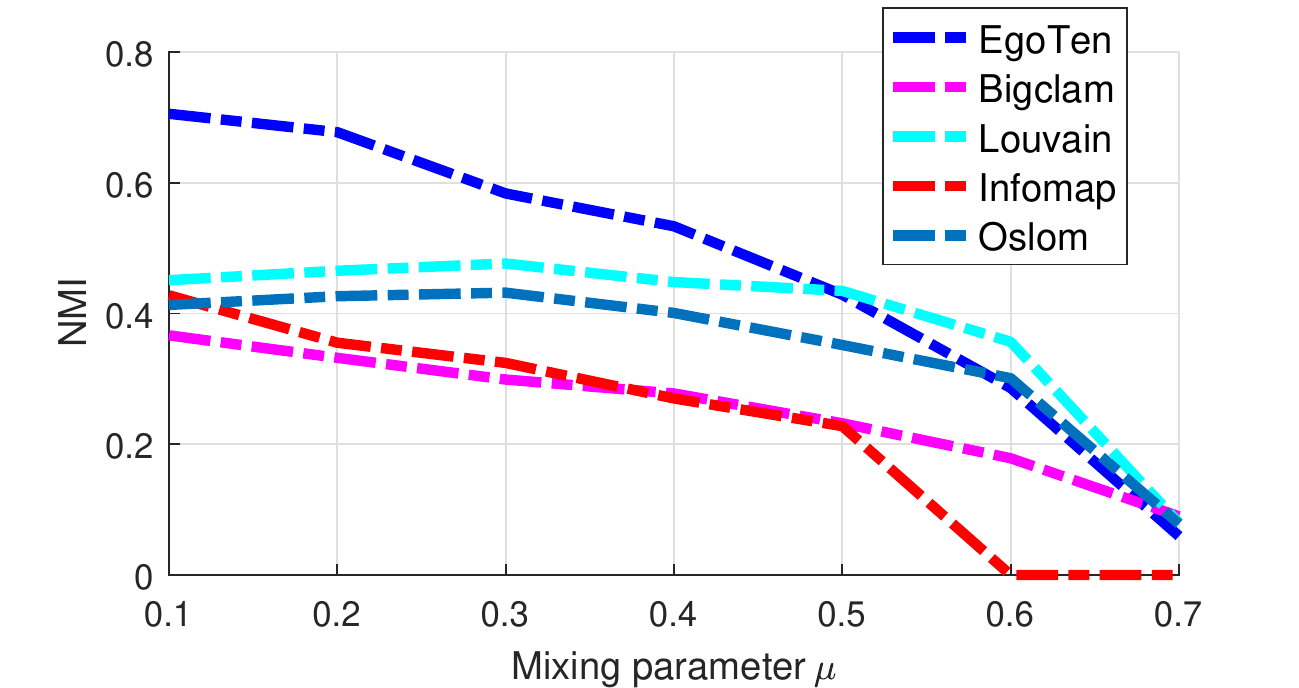}}
		\centerline{(a)   }\medskip
	\end{minipage}
	\vspace{-0.0cm}
	\begin{minipage}{1\linewidth}
		\centering
		\vspace{0.0cm}
		\centerline{\includegraphics[width=01\textwidth]{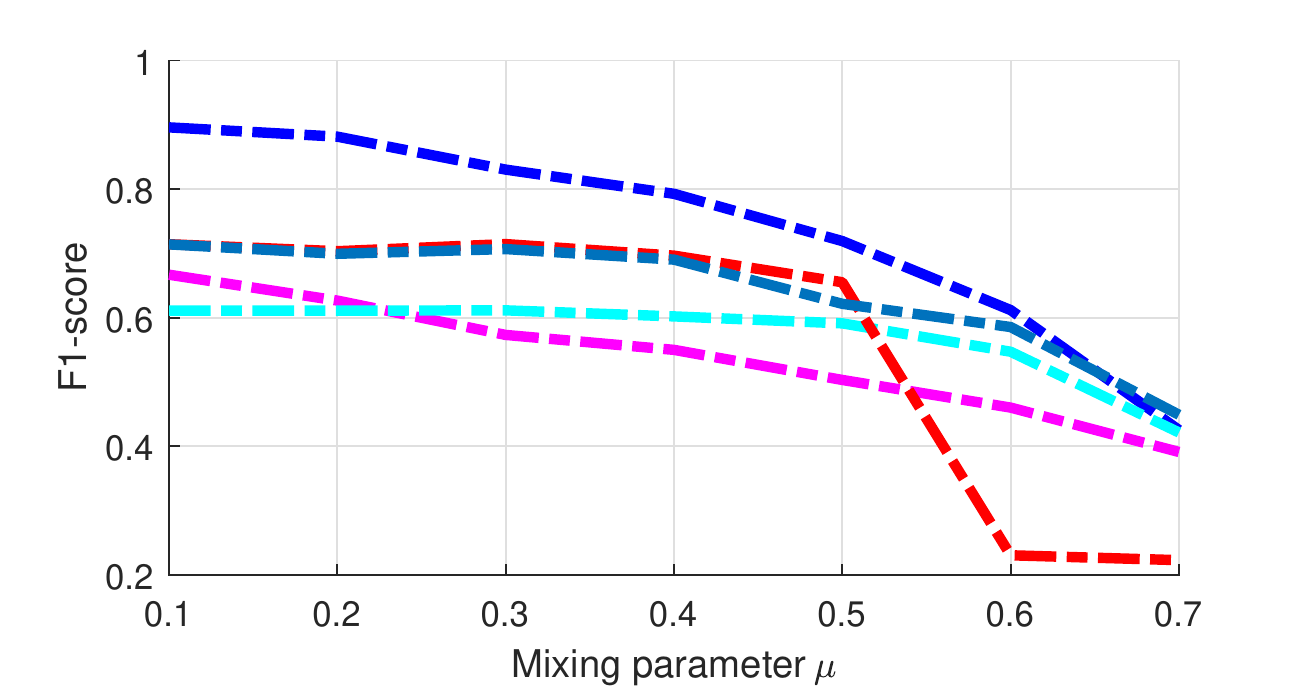}}
		\centerline{(b) }\medskip
	\end{minipage}
	\hfill
	\caption{Performance of  EgoTen and state-of-the-art algorithms in terms of (a) NMI; and, (b) average F-1 score, for LFR networks of $N=1,000$, $o_n=300$, and $o_m=5$  versus  $\mu$.}
	\label{fig:res}
\end{figure}

Performance  is reported in terms of NMI, and F1-score. As Figures 8 and 9 corroborate,  EgoTen offers the highest performance for a wide range of  $\mu$, as well as  $o_n$, thanks to the  reinforced structure  of the egonet-tensor. 

\begin{figure}[h!]
	\begin{minipage}{1\linewidth}
		\centering
		\vspace{-0.0cm}
		\centerline{\includegraphics[width=1\textwidth]{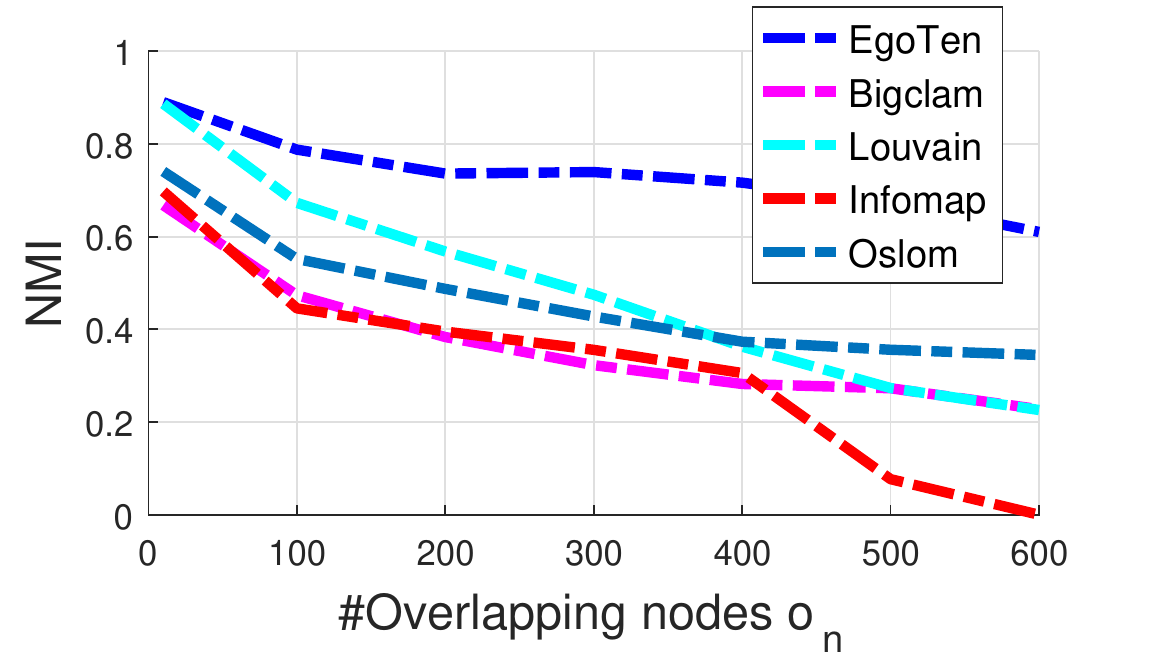}}
		\vspace{-0.0cm}
		\centerline{(a)   \vspace{-0.0cm}}\medskip
	\end{minipage}
	\vspace{-0.0cm}
	\begin{minipage}{1\linewidth}
		\centering
		
		\centerline{\includegraphics[width=01\textwidth]{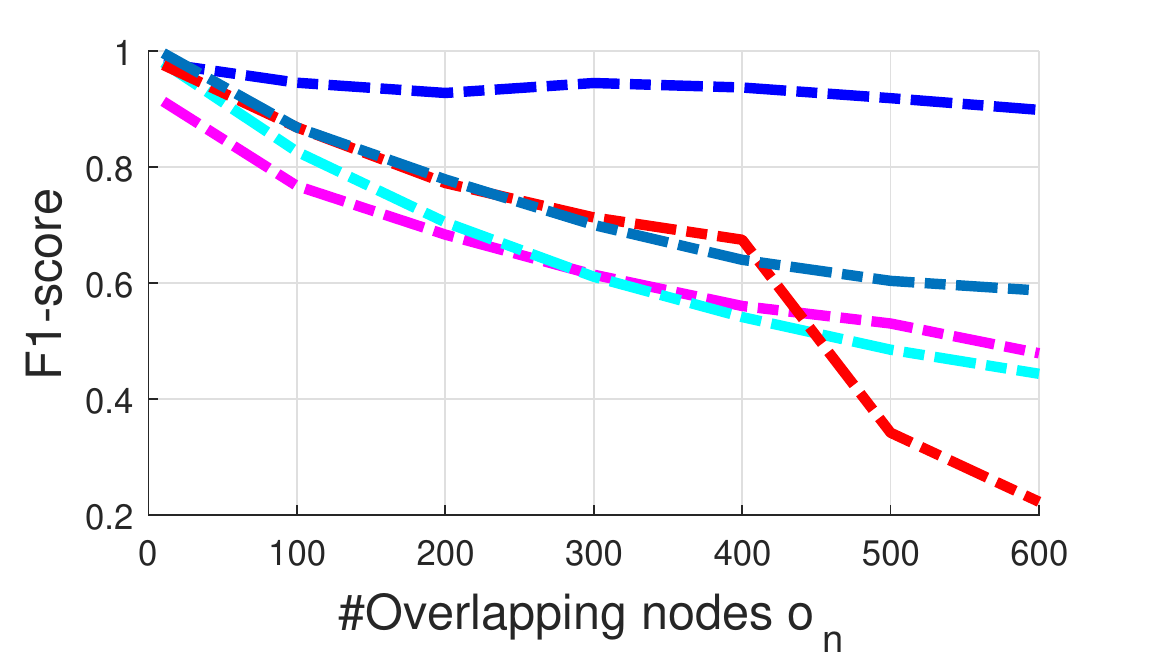}}
		\centerline{(b) }\medskip
	\end{minipage}
	\hfill
	\caption{Performance of  EgoTen and state-of-the-art algorithms in terms of (a) NMI; and, (b) average F1-score, for LFR networks of $N=1,000$, $\mu=0.2$, and	$o_m=5$  versus  $o_n$.}
	\label{fig:res}
\end{figure}

\begin{figure}[h!]\label{fig:complexity}
	\begin{minipage}{1\linewidth}
		\centering
		\vspace{-0.0cm}
		\centerline{\includegraphics[width=1\textwidth]{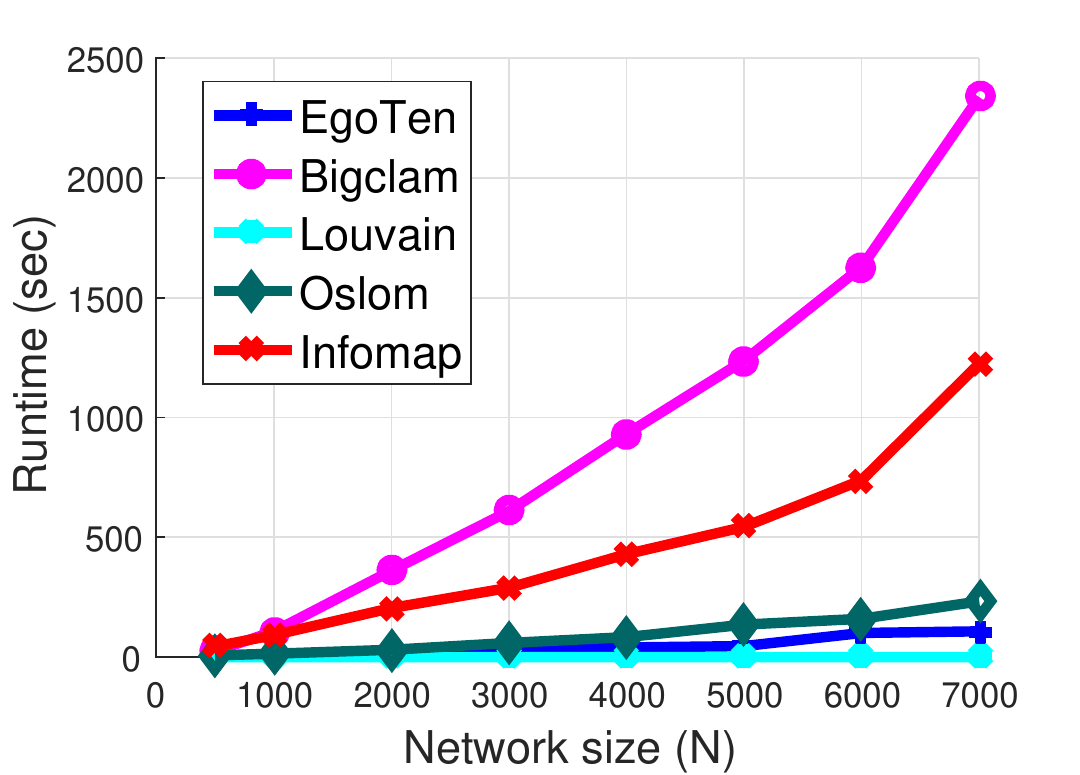}}
		\vspace{-0.0cm}
	\end{minipage}
	\vspace{-0.cm}
	\caption{Scalability of different algorithms as $N$ grows.}
	\label{fig:res}
\end{figure}

Regarding the scalability of the proposed EgoTen algorithm, Fig. 10 depicts run time versus network size $N$ while average nodal degree is kept constant at $\bar{d}=100$. As the plot corroborates, exploitation of the sparsity in the egonet-tensor provides the algorithm with scalability, while the same can not be claimed for all other competitors. %

\subsection{Time-varying graphs}
In this subsection, the performance of the proposed EgoTen in Alg. 4 for community identification over time-varying networks is assessed. To this end, we have generated a synthetic network with $N=1,000$ nodes for a span of $T=20$ slots. Initially at $t=1$, the networks is generated with two distinct communities, each containing of $500$ nodes. For $t>1$, the community association of $600$ randomly selected  nodes remains unchanged, whereas the other $400$ nodes migrate from their original community to a third newly-formed community. Transition slot $\tau_n$ for each of these nodes is identically drawn from a normal distribution $\mathcal{N}(10,1)$. For any time slot $t$, the network edges are  drawn according to a block stochastic model, where nodes within the same community are connected with probability $0.3$, and out-of-community edges are drawn with probability $0.1$. EgoTen's performance is compared with that of constrained NMF, for which  $\bU$ and $\bV$ per $t$ is used as initialization for $t+1$ to provide NMF with consistency across time.  

Performance  is measured in terms of NMI, and it is averaged over $20$ realizations of the network in Fig. 12. Furthermore, Fig. 13 illustrates the identified communities for different nodes  across time for a realization. That is, for any $t$ on the x-axis, nodes associated with the same community are shown with the same color.  The plot depicts ground truth as well as EgoTen and NMF results, where we have perturbed ordering of nodes for a better visualization. Clearly,  EgoTen successfully identifies the two initial communities, as well as three communities after the migration of a subset of nodes,   presenting  solid blocks similar to those  in the ground truth, while communities detected via constrained NMF are of lower quality.

\begin{figure}	
		\centering
		\centerline{\includegraphics[width=0.5\textwidth]{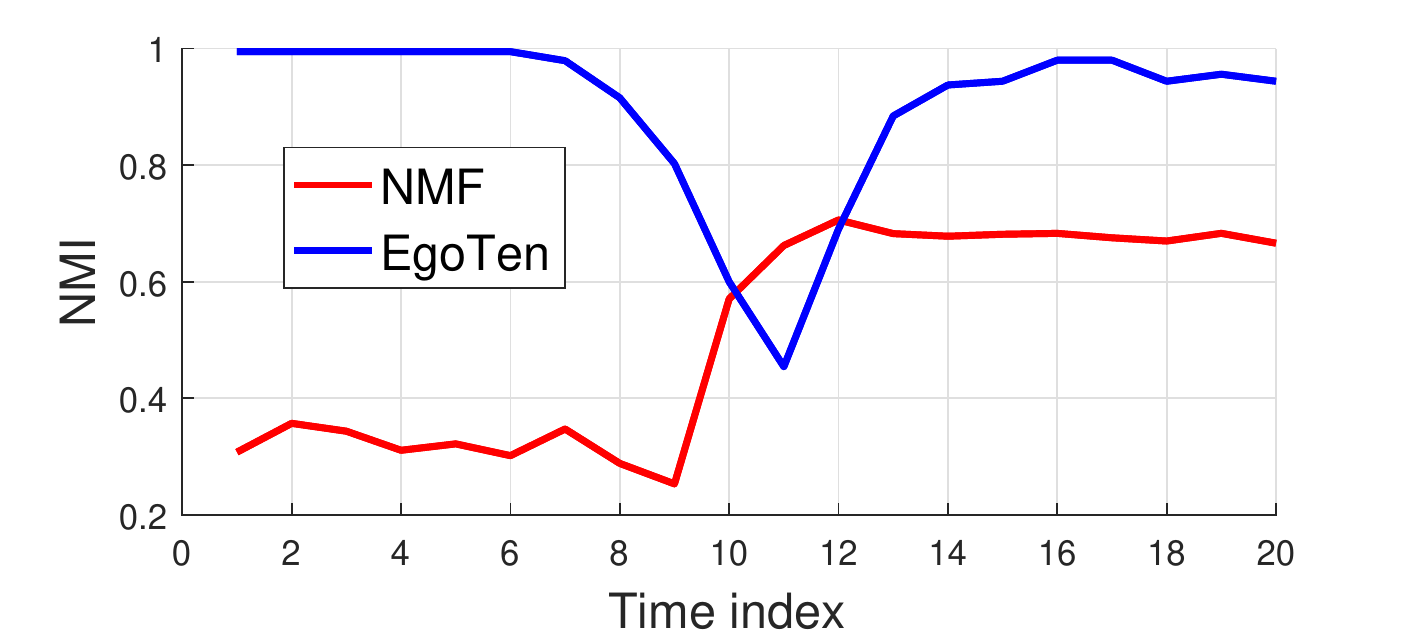}}
		\caption{NMI for the detected communities across  time for  synthetic time-varying graphs.}
	\label{fig:res}
\end{figure}

\begin{figure}	
		\centering
		\centerline{\includegraphics[width=0.5\textwidth]{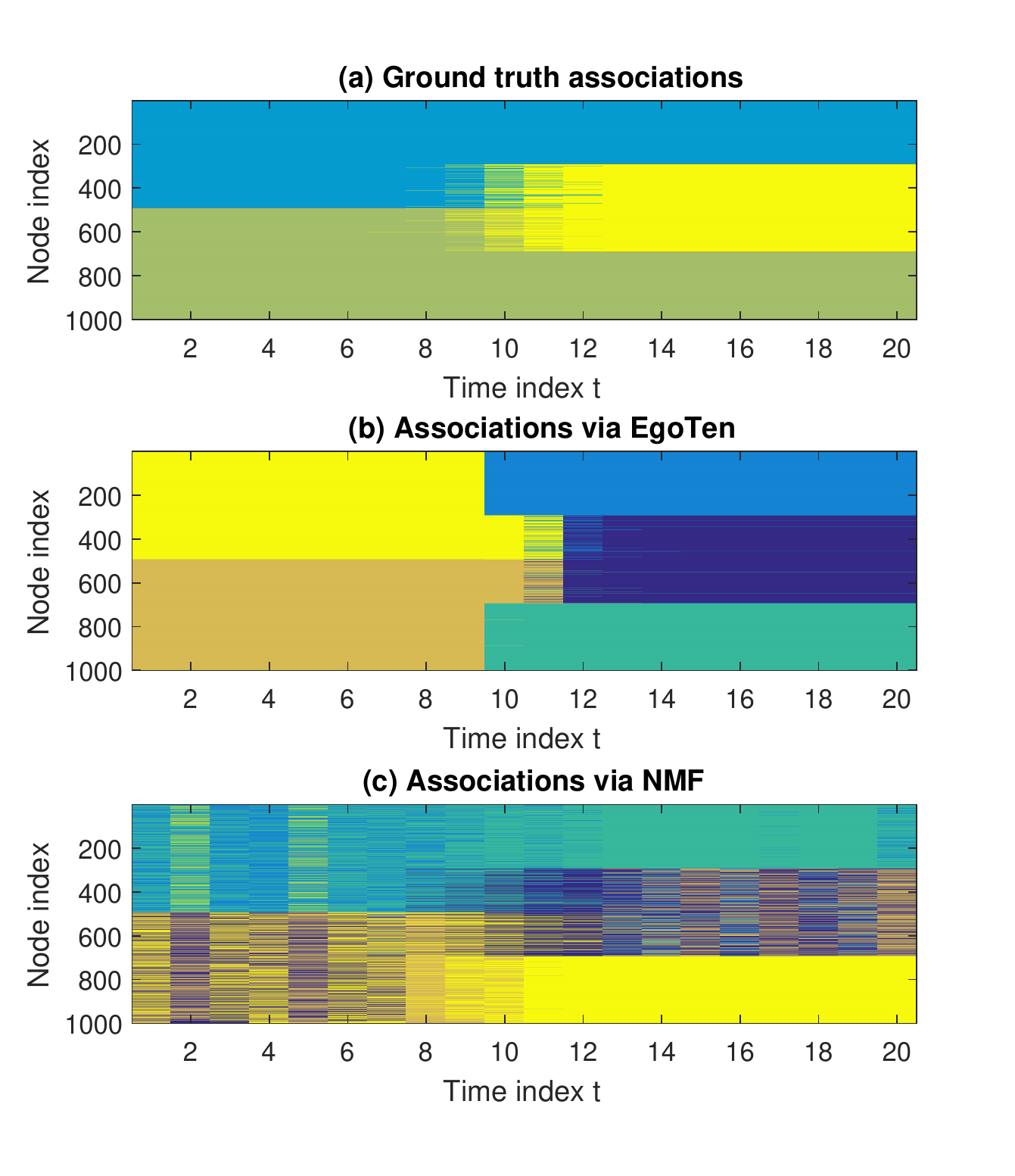}}
		\caption{Time-varying community association of nodes, where same communities are color-coded similarly; thus, for a given  $t$ on the x-axis, nodes in a community have the  same color in (a) ground truth; and results via (b) EgoTen; and, (c) NMF.}
	\label{fig:res}
\end{figure}

\subsection{Real world Networks }
In this section EgoTen is utilized for performing community detection on a number of real-world networks, tabulated in Table {\ref{tab:datasets}}. To study the quality of detected communities for real-world networks -whose ground truth community association is often unavailable- we examine the conductance of detected communities. In particular, given a cover $\mathcal{C} = \{\mathcal{C}_1, ..., \mathcal{C}_K\}$, let us compute the conductance $\phi(\mathcal{C}_i)$ for $i=1,\cdots, K$. Corresponding to a value $\nu \in [0 \; 1]$, let us define the set of communities whose conductance is less than $\nu$, i.e.,  $\mathcal{S_\nu } :=  \{ \mathcal{C}_j|\phi( \mathcal{C}_j)<\nu \}$. Then, coverage($\nu$) is defined as 
\begin{equation}
\text{coverage}(\nu):=
\dfrac{\Big|  \underset{ \mathcal{C}_i \in \mathcal{S}_\nu} \cup \mathcal{C}_i\Big|}{N}
\end{equation}
Consequently, conductance-coverage curve is plotted by varying the value $\nu$ from 0 to 1 on the $y$-axis and reporting the corresponding coverage value on the $x$-axis. As low values of conductance correspond to more cohesive communities, smaller area under curve (AUC) implies better performance. Fig. 12 plots the coverage-conductance curve and Table \ref{tab:auc} tabulates AUC as well as average conductance defined in \eqref{average_conductance}. Since Louvain does not allow for overlapping nodes, the two metrics coincide, hence the corresponding one column  in Tabel II. As the results corroborate, the quality of detected communities via  EgoTen is closely competing or outperforming the ones provided by other methods, while it remains robust to ``resolution limit''~\cite{fortunato2007resolution} observed in Oslom in Fig. 12 (a) and Infomap in Fig. 12 (d) whose   performance are limited to detecting only large communities.

\begin{table}[ht]
	\centering
	\caption{Real-world networks.}
	\begin{tabular}{||c |c c  ||} 
		
		\hline
		$r_a$ &    Size $N$ &  Edges\\	
		\hline 
		\hline
		{Dolphins} & 62	 &  159  \\
		\hline
		
		{Les miserable} 	& 77  &  254\\
		\hline
		{Football} & 115	 & 613  \\
		\hline
		
		{Facebook}   & 4039&    88,234 \\
		\hline
		
	\end{tabular}
	
	\label{tab:datasets}
\end{table}

\begin{figure}[ht]
	\includegraphics[scale=0.45]{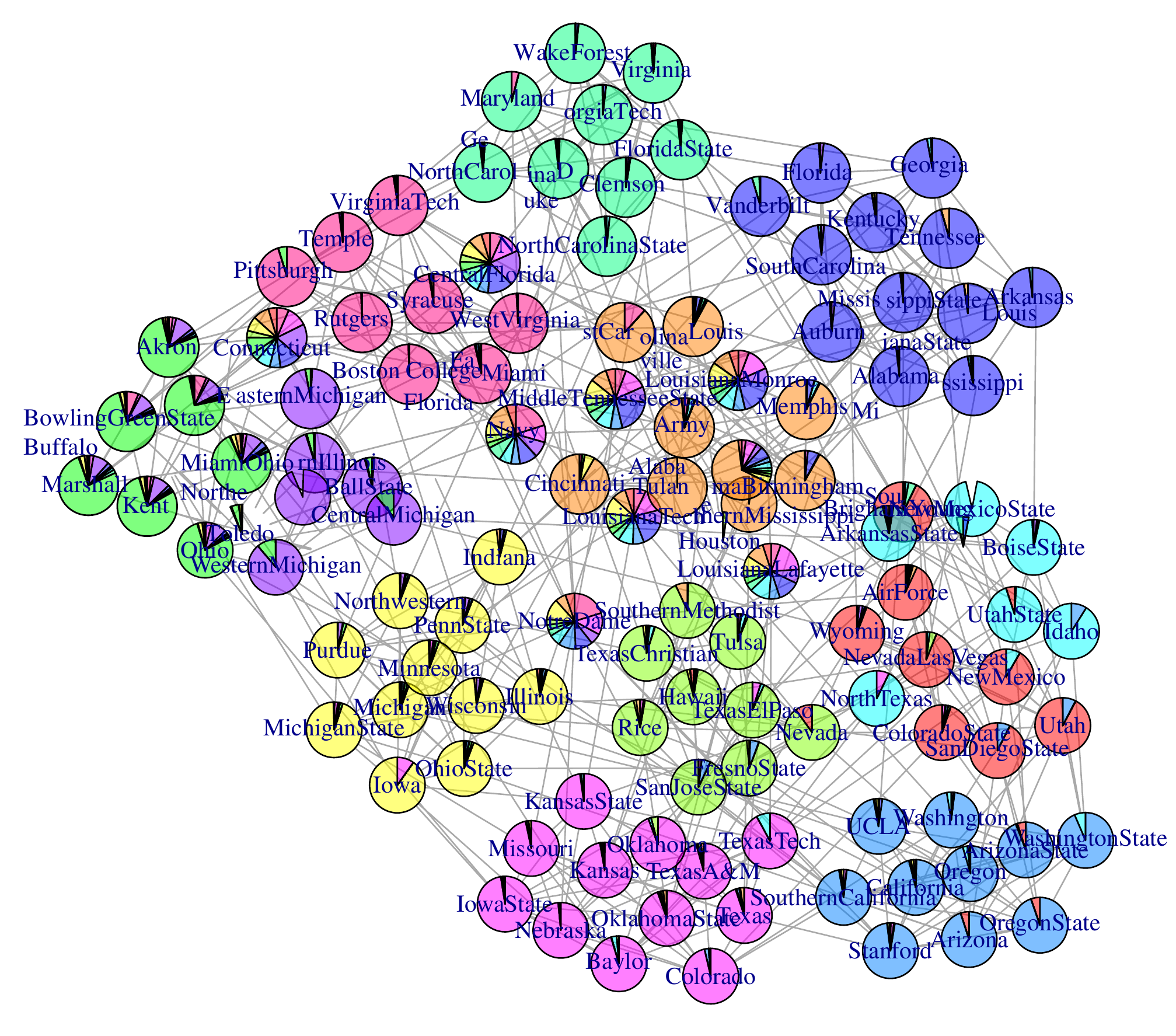}
	\vspace{-0.1cm}
	\caption{Visualization of the American College Football Network with $N=115$ and $K=12$.  Different colors correspond to different detected communities, and the pie-charts reveal soft community association of the nodes.}
\end{figure}

\begin{table*}[h!]
	\caption{Quality of detected cover on real-world networks in terms of AUC and average conductance.}
	\begin{tabular} {|c|| c c   c|| c  c c|| c   c|| c  c c|| c c  c||}
		\hline
 &    \multicolumn{3}{c||}{ EgoTen} & \multicolumn{3}{c||}{Bigclam } &\multicolumn{2}{c||}{ Louvain } & \multicolumn{3}{c||}{Infomap}   & \multicolumn{3}{c||}{Oslom  } \\	
		&  $|\mathcal{C}|$ & AUC	 & $\bar{\phi}({\mathcal{C}})$ &  $|\mathcal{C}|$ & AUC	 & $\bar{\phi}({\mathcal{C}})$ &   $|\mathcal{C}|$ & AUC	  (or $\bar{\phi}({\mathcal{C}})$)  &  $|\mathcal{C}|$ & AUC	 & $\bar{\phi}({\mathcal{C}})$ &     $|\mathcal{C}|$ & AUC	 & $\bar{\phi}({\mathcal{C}})$ 	  \\
		\hline 
		{Dolphins} & 10	 &  0.2984 & 0.4584&  5 &  0.6176 &  0.6176   	 & 11  & 0.3902& 10 & 0.3201 &  0.4012& 2 & 0.1034 &  0.1034 \\
		
		{Les miserable} 	& 5 &  0.2803&   0.2803  &  10&   0.6042 &     0.4666& 15 & 0.3343& 12 & 0.3534 &  0.4052& 3 & 0.2127 &  0.2182 \\
		
		{Football} & 15	 &  0.4085&  0.3480 &    15	&  0.4989 & 0.4101 & 15 & 0.3752 &  12& 0.3974 & 0.3468& 11 & 0.3430 & 0.3037\\
		
		{Facebook}   & 100&  0.3768    & 0.4931  &100  & 0.3798 & 0.4973  &100 &  0.1329&5 & 0.0370 & 0.0360& 110 & 0.4998 & 0.4955 \\
		\hline
		
	\end{tabular}
	\label{tab:auc}
\end{table*}

\begin{figure*}[h!]
	
	\begin{minipage}[b]{.48\linewidth}
		\centering
		\centerline{\includegraphics[width=1\textwidth]{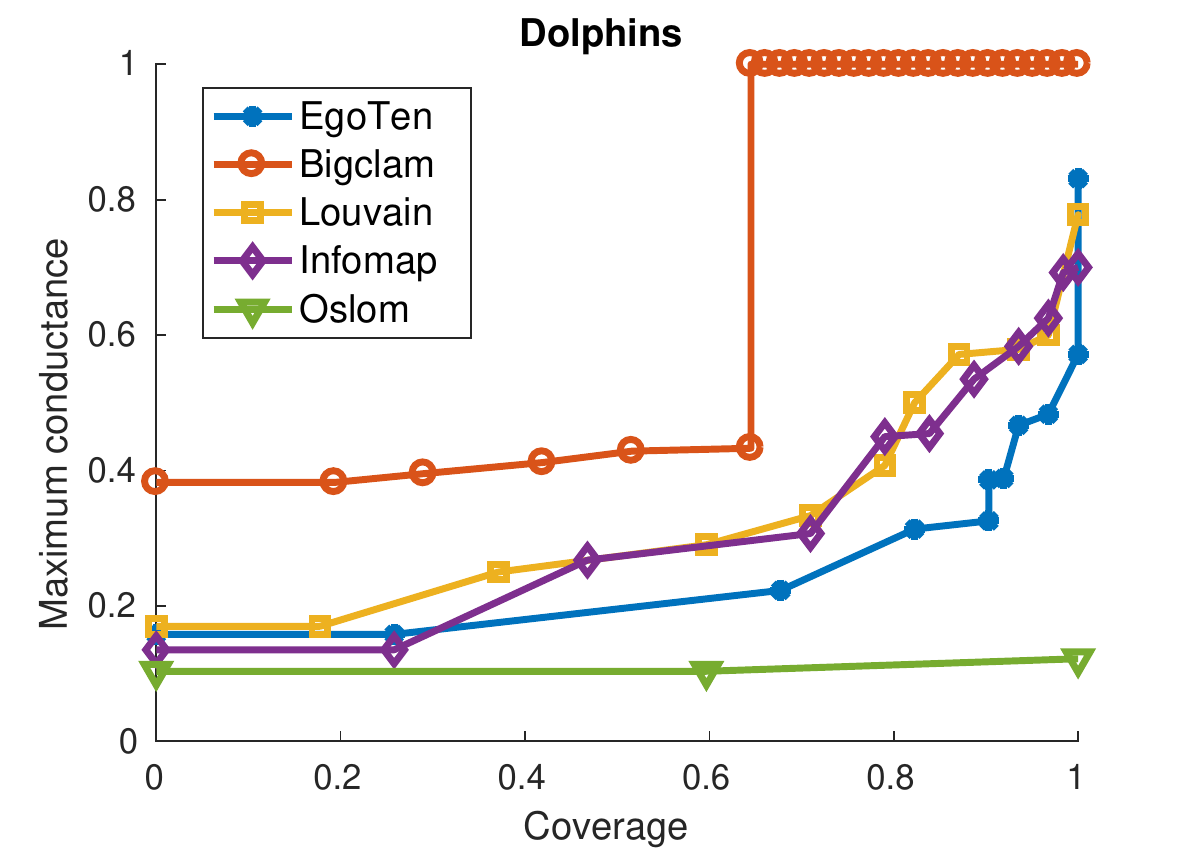}}
		\centerline{(a) }\medskip
	\end{minipage}
	\begin{minipage}[b]{.48\linewidth}
		\centering
		\centerline{\includegraphics[width=1\textwidth]{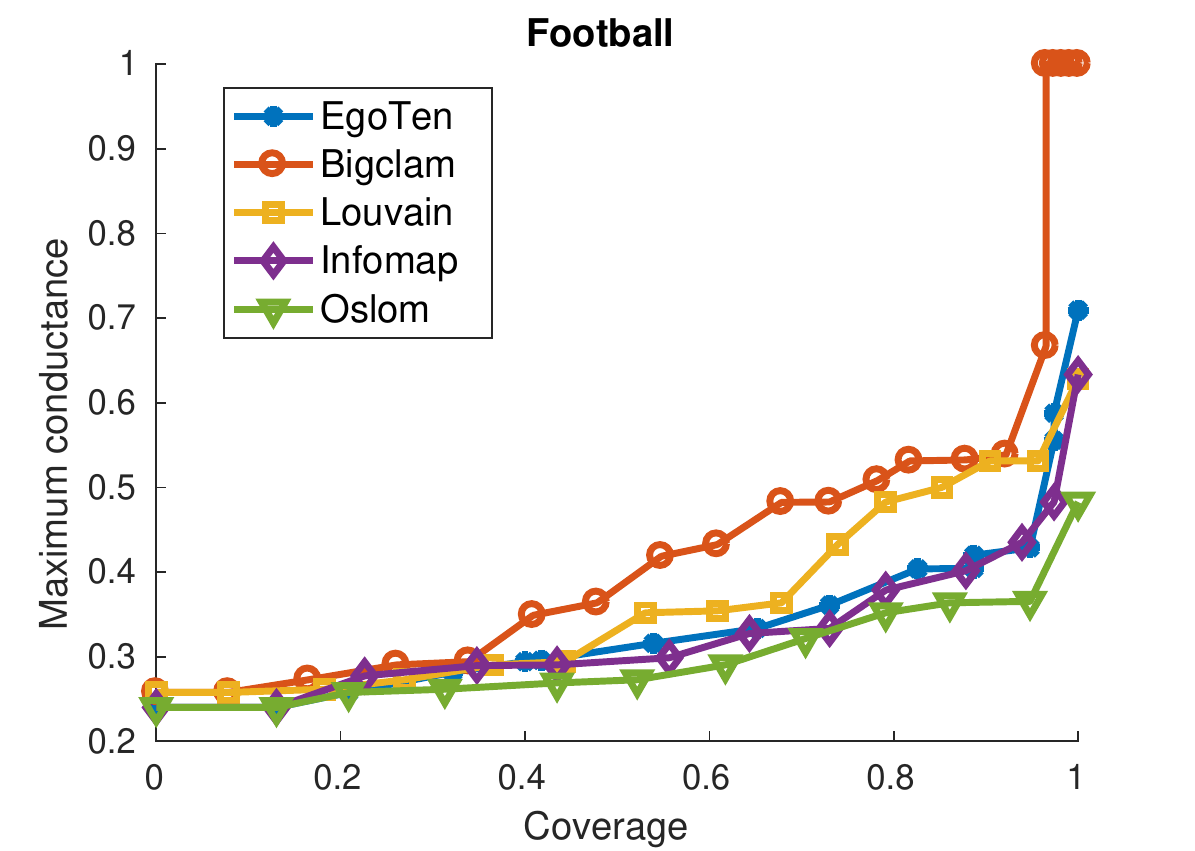}}
		\centerline{(b) }\medskip
	\end{minipage}
	\begin{minipage}[b]{.5\linewidth}
		\centering
		\centerline{\includegraphics[width=1\textwidth]{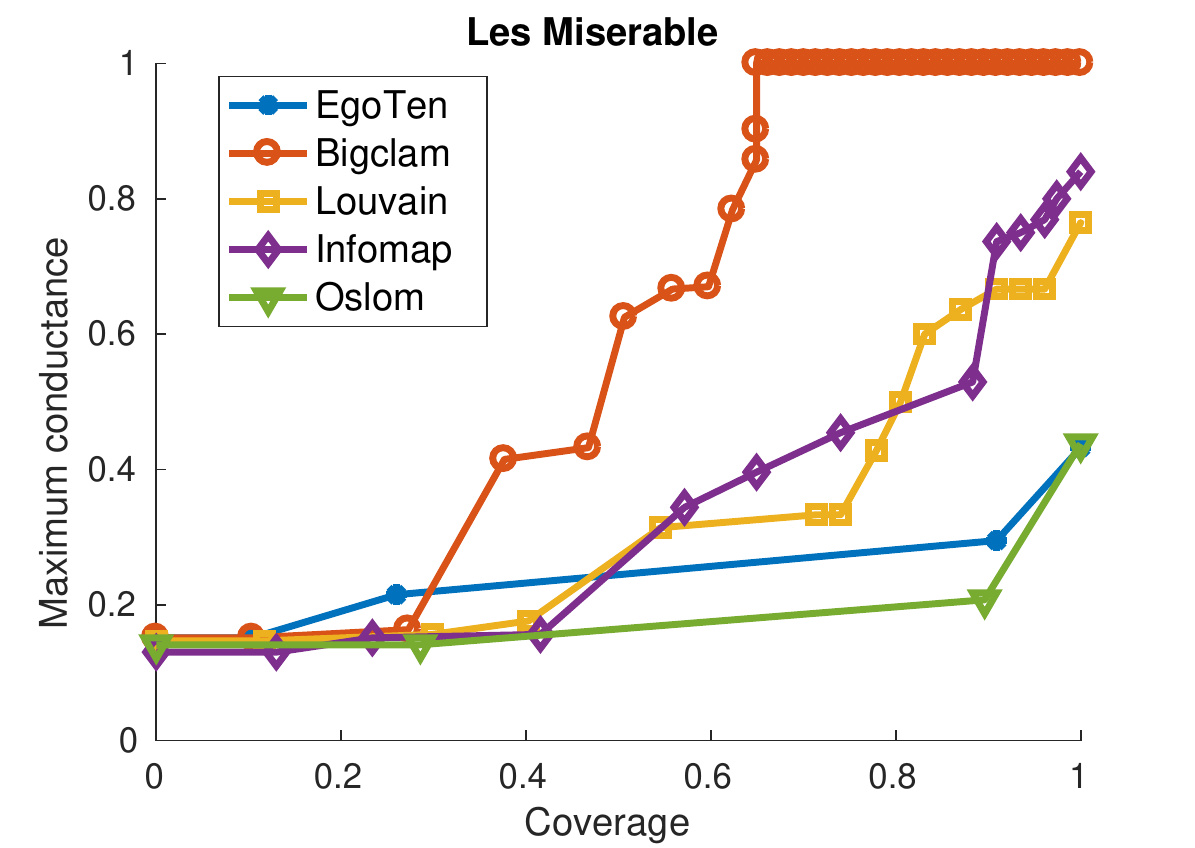}}
		\centerline{(c) }\medskip
	\end{minipage}
	\begin{minipage}[b]{0.5\linewidth}
		\centering
		\centerline{\includegraphics[width=0.9\textwidth]{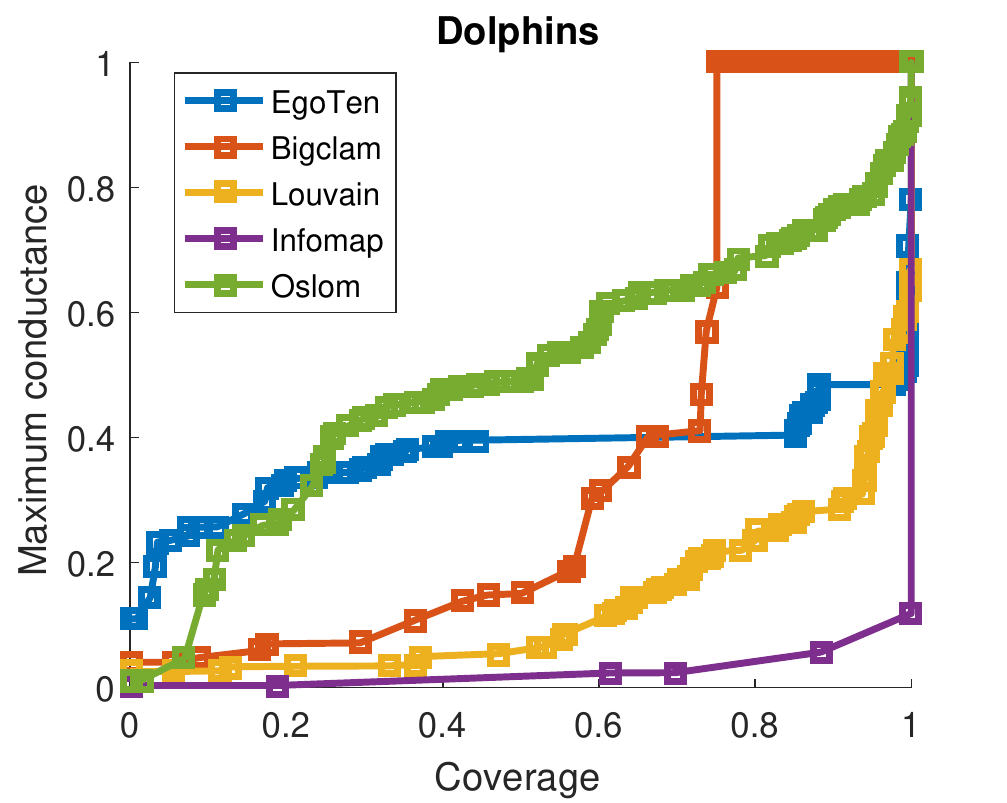}}
		\centerline{(d) }\medskip
	\end{minipage}
	\caption{Maximum-conductance versus coverage for real-world networks with underlying community structure. Lower curves correspond to better performance.}
\end{figure*}

\section{Conclusion and Remarks }\label{sec:conclusion}
By viewing  networks as the union of nodal egonets, a novel tensor-based representation for  capturing high-order nodal connectivities has been introduced.  The induced redundancy in the constructed egonet-tensor bestows the novel representation with rich structure, and is  utilized for community detection by casting the problem as a constrained tensor decomposition task. Utilization of tensor sparsity as well as parallel computation endow the algorithm with scalability, while the structured redundancy enhances the performance against overlapping and highly-mixed communities.  The proposed framework is broadened to accommodate time-varying graphs, where a four-dimensional tensor enables simultaneous community identification over the entire horizon yielding an improved performance.

As a natural extension, one can generalize the tensor-based representation to account for adjacency matrices capturing the connectivity of  $2$, $3, \ldots, d_{\max}$-hop neighbors. This approach indeed highlights the tradeoff between flexibility and redundancy, as memory and computational intensity of the corresponding CPD as well as proper tuning of parameters will influence the quality of the detected communities. One could analyze this tradeoff to further characterize how the quality of detected communities evolves as the  coverage of \textit{extended-egonets} increases; however, this goes beyond the scope of this work, and is left for future investigation.

\bibliographystyle{IEEEtran}

\bibliography{ref_commID}

\end{document}